\documentclass[fleqn,twoside]{article}
\usepackage{espcrc2}
\usepackage{natbib,psfig}
\usepackage[dvips]{graphicx}
\usepackage{amsbsy}
\usepackage{epsfig}
\usepackage{psfrag}
\def \bea{\begin{eqnarray}}
\def \beq{\begin{equation}}
\def \eea{\end{eqnarray}}
\def \eeq{\end{equation}}
\def\lesssim{\mathrel{\hbox{\rlap{\hbox{\lower4pt\hbox{$\sim$}}}\hbox{$<$}}}}
\def\gtrsim{\mathrel{\hbox{\rlap{\hbox{\lower4pt\hbox{$\sim$}}}\hbox{$>$}}}}

\renewcommand{\cite}{\nocite}

\title{Prospects for radio detection of ultra-high energy cosmic rays and neutrinos}

\author{
H. Falcke\address{ASTRON, P.O. Box 27990 AA Dwingeloo,
The Netherlands, falcke@astron.nl},
P. Gorham\address{Department of Physics \& Astronomy,  Univ. of Hawaii at Manoa,
Honolulu, HI, 96822, USA, gorham@phys.hawaii.edu},
R.J. Protheroe\address{Department of Physics, University of Adelaide, 
SA 5006, Australia, rprother@physics.adelaide.edu.au}}

\begin{document}

\begin{abstract}

The origin and nature of the highest energy cosmic ray events is
currently the subject of intense investigation by giant air shower
arrays and fluorescent detectors.  These particles reach energies well
beyond what can be achieved in ground-based particle accelerators and
hence they are fundamental probes for particle physics as well as
astrophysics. One of the main topics today focuses on the high energy
end of the spectrum and the potential for the production of
high-energy neutrinos.  Above about $10^{20}$~eV cosmic rays from
extragalactic sources are expected to be severely attenuated by pion
photoproduction interactions with photons of the cosmic microwave
background.  Investigating the shape of the cosmic ray spectrum near
this predicted cut-off will be very important. In addition, a
significant high-energy neutrino background is naturally expected as
part of the pion decay chain which also contains much information.

Because of the scarcity of these high-energy particles, larger and
larger ground-based detectors have been built. The new generation of
digital radio telescopes may play an important role in this, if
properly designed. Radio detection of cosmic ray showers has a long
history but was abandoned in the 1970's.  Recent experimental
developments together with sophisticated air shower simulations
incorporating radio emission give a clearer understanding of the
relationship between the air shower parameters and the radio signal,
and have led to resurgence in its use.  Observations of air showers by
the SKA could, because of its large collecting area, contribute
significantly to measuring the cosmic ray spectrum at the highest
energies.  Because of the large surface area of the moon, and the
expected excellent angular resolution of the SKA, using the SKA to
detect radio Cherenkov emission from neutrino-induced cascades in
lunar regolith will be potentially the most important technique for
investigating cosmic ray origin at energies above the photoproduction
cut-off.

\vspace{1pc}

\end{abstract}

%\begin{keyword}
%cosmic rays, neutrinos, radio detection
%\end{keyword}
%\end{frontmatter}
\maketitle
%\chapter{Ultra-High Energy Cosmic Rays and Neutrinos}
\section{Observational and theoretical motivation}

Understanding the origin of the ultra-high energy cosmic rays
(UHECR), the highest energy particles observed in nature,
is of great importance as it may impact our understanding of particle physics, 
fundamental cosmology, and extremely energetic phenomena in the
Universe.  The energy spectrum of UHECR extends up to at least
$10^{11}$~GeV, and in the rest frame of a UHECR proton, photons of
the 2.73~K cosmic microwave background radiation (CMBR) are
strongly blue-shifted to gamma-ray energies.  The threshold for
Bethe-Heitler pair production and pion photoproduction by UHECR
protons on the CMBR are close to $2 \times 10^{8}$~GeV and $2
\times 10^{10}$~GeV, such that protons at $3\times$$10^{10}$~GeV
and 3$\times$$10^{11}$~GeV typically lose a large fraction of their
energy in a time of 1~Gpc/c ($3 \times 10^9$y) and 10~Mpc/c ($3 \times
10^7$y), respectively.  This would imply that sources of ultra-high
energy cosmic rays would have to be close if the particles themselves
behave as predicted. The importance of pion photoproduction on the
CMBR was first noted by Greisen (1966) \cite{greisen} and Zatsepin \&
Kuzmin (1966) \cite{zatsepin} and the cut-off they predicted at $\sim
10^{11}$~GeV is referred to as the ``GZK cut-off''.

Yet, until today neither the nature of the particles nor of their
accelerators has been revealed. It is well established that in some
astrophysical magnetized plasma regions particles (leptons) are
accelerated, but whether and where this holds for UHECR is
unclear. Some basic constraints can nonetheless be given.  Cosmic ray
acceleration sites must be large enough to contain the gyroradius of
the accelerated particles, as well as having scattering centres with
appropriate velocities.  In addition, the acceleration must be
sufficiently rapid that high energies can be achieved in an
accelerator's lifetime, and that energy losses by pion photoproduction
and synchrotron radiation do not cut off the spectrum too soon (Hillas
1984\cite{Hillas84}).  It is currently unknown whether the UHECR are
Galactic or extragalactic in origin.  Composition measurements are
also important because if UHECR are observed to include nuclei other
than protons then these must be from Galactic, or very nearby
extragalactic acceleration sources to avoid photodisintegration (see
e.g.~Yamamoto et al. 2004\cite{Yamamoto}).  However, the promising
extragalactic source candidates for UHECR above $10^{11}$~GeV are
typically at distances too far for UHECR to reach us unaffected by
interactions with the CMBR. This is the basic dilemma we are faced
with today.

\subsection{UHECR observations}

\begin{figure}[htb] 
\centerline{\epsfig{file=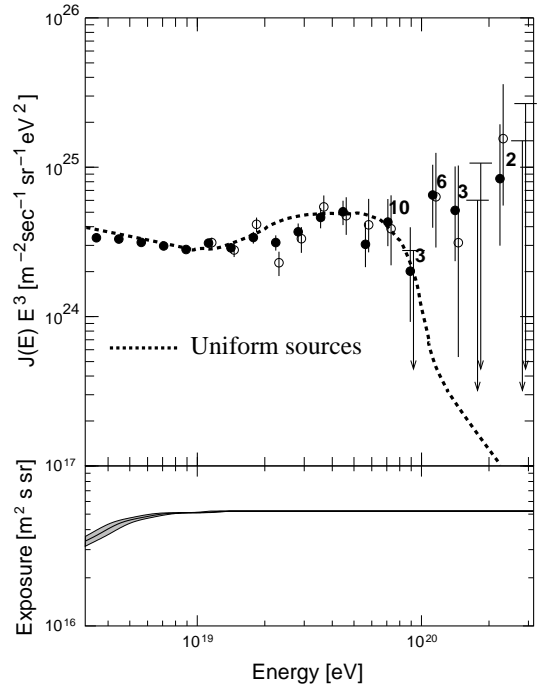,width=0.48\textwidth}}
%%-%\vspace*{10pt}
\caption{Spectrum of UHECR detected by AGASA. Numbers attached to points show the number of events in each energy bin.  (From Takeda et al. 2003\protect\cite{rf:TAK}.)
\label{agasa}
}
\end{figure}

% \begin{figure*}[htb] 
% \centerline{\epsfig{file=AGASA_arrivalsEQU_1960_A20.eps,width=85mm}}
% %-%\vspace*{10pt}
% \caption{  (From Stan at al.\ \protect
% \cite{Stanevatal200}.)
% \label{power}
% }
% \end{figure*}

Below the GZK cut-off UHECR may, to some extent, point back to
their sources depending on the structure and strength of the
magnetic field between the sources and our Galaxy.  No
statistically compelling anisotropy has been detected in the UHE
CR.  The energy spectrum of UHECR detected by AGASA is shown in
Fig.~\ref{agasa}.  There are two main problems at present: the
flux of UHECR is so low that few events have been detected for
reliable conclusions concerning the presence or absence of a GZK
cut-off or any anisotropy, and in the case of a GZK
cut-off spectral information above the cut-off would be lost.
New experiments such as HiRes and the 3000 km$^2$ Peirre Auger
Observatory will help to address the the first issue and, because
of its huge area, use of the SKA may also help here by direct
radio detection of UHECR air showers.  The second problem, that
of loss of spectral information above the GZK cut-off, is best
explored through UHE neutrino astronomy and again the use of the
SKA, either to detect neutrino-induced air showers or radio
Cherenkov bursts from electromagnetic cascades in lunar regolith
initiated by interactions of UHE neutrinos, has the potential to
greatly add to our knowledge of the the origin of the highest
particles in nature.

\subsection{The GZK problem}

\begin{figure}[htb!] 
\centerline{\epsfig{file=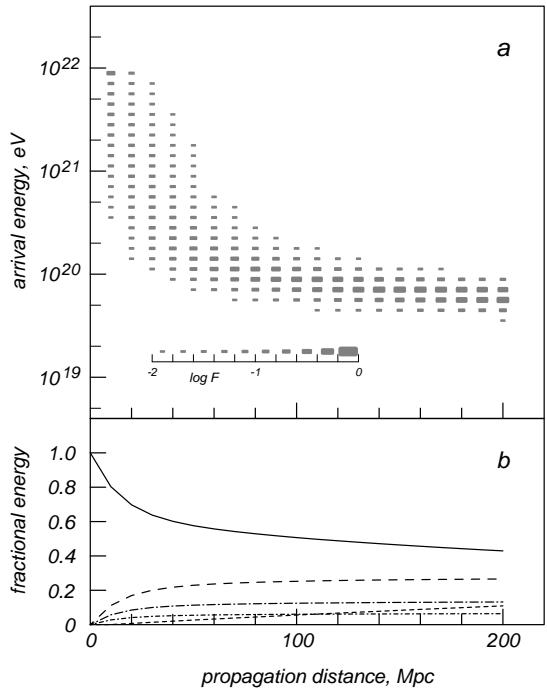,width=0.48\textwidth}}
%-%\vspace*{10pt}
\caption{ (a) Arrival energy distribution for protons injected
with energy between 10$^{21.9}$ and 10$^{22}$~eV after
propagation on 10, 20, ... 200 Mpc.  (b) Fractional energy
contained in nucleons (solid line), $\gamma$--rays from
photoproduction (long dashes) and BH pair production (short
dashes) for protons injected with a $E^{-2}$ power law spectrum
with an exponential cutoff at 10$^{21.5}$~eV. The dash--dot lines
show the fractional energy in muon (long) and electron (short)
neutrinos and antineutrinos.  (From Stanev at al.\ \protect
\cite{Stanevatal2000}2000.)
\label{power}
}
\end{figure}

Due to interactions with the CMBR, there is expected to be a
spectral downturn, the GZK cut-off, for particles which have
travelled more than a few tens of Mpc.  Figure~\ref{power} shows
the distribution in energy as a function of distance travelled
for UHECR protons with initial energies close to $10^{22}$~eV.  As
can be seen, after 80~Mpc no protons have energies above $3
\times 10^{20}$~eV.  Of course, UHECR protons are also deflected
by extragalactic magnetic fields, and so any source of $3 \times
10^{20}$~eV UHECR would need to be much nearer than 80~Mpc.
However, several experiments have reported CR events with
energies above $10^{20}$~eV with the highest energy event having
$3 \times 10^{20}$~eV (Bird et al.\cite{rf:FE3}1995).  Very
recent data from the two largest aperture high energy cosmic ray
detectors are contradictory: AGASA (Takeda et al.\cite{rf:TAK} 2003) observes no GZK cut-off while HiRes
(Abbasi et al.\cite{HiRes04} 2004) observes a cut-off
consistent with that expected.  A systematic over-estimation of
energy of about 25\% by AGASA or under-estimation of energy of
about 25\% by HiRes could account the discrepancy (Abbasi et
al.\cite{HiRes04} 2004), but the continuation of the UHECR
spectrum to energies well above $10^{20}$~eV is now far from
certain.  Future measurements with Auger (Auger Collaboration
2001) should resolve this question.  Whether or not the spectrum
does extend well beyond $10^{20}$~eV, determining the origin of
these particles could have important implications for
astrophysics, cosmology and particle physics.

\subsection{The acceleration problem}

\begin{figure}[htb!]
\centerline{~~~~~~~~\epsfig{file=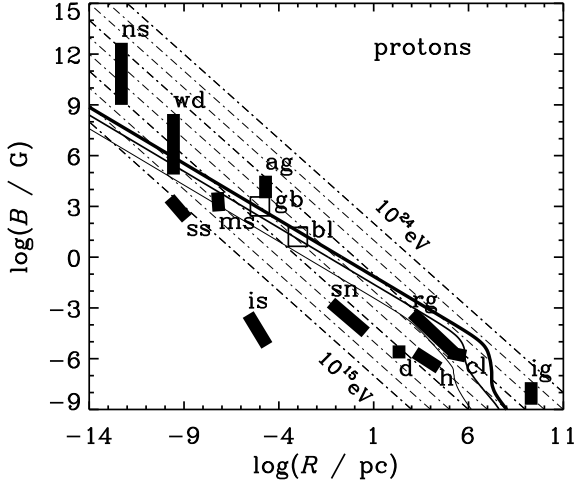,width=0.63\textwidth}}
\caption{``Hillas plot'' showing (chain curves) magnetic field
vs.\ gyroradius for proton momenta $10^{15}$, $10^{16}$, \dots,
$10^{24}$~eV/c.  The solid curves bound the parameter space of
accelerated particles for a given acceleration rate parameter
(see text). Typical size and magnetic field of possible
acceleration sites (taken from Hillas \protect\cite{Hillas84}1984)
are shown for neutron stars (ns), white dwarfs (wd), sunspots
(ss), magnetic stars (ms), AGN (ag), interstellar space (is),
supernova remnants (sn), radio galaxy lobes (rg), galactic disk
(d) and halo (h), clusters of galaxies (cl) and intergalactic
medium (ig).  Typical jet-frame parameters of the synchrotron
proton blazar model (M\"uecke et al.\protect\cite{Mueckeetal03} 2003)
and gamma ray burst model (Pelletier \&
Kersate\protect\cite{PelletierKersate00} 2000) are indicated by the
open squares labelled ``bl'' and and ``gb''.  (From
Protheroe~\protect\cite{Protheroe2004}2004.)}
\label{proton_emax}
\end{figure}

By plotting magnetic field vs.\ size of various astrophysical
objects (Fig.~\ref{proton_emax}), Hillas (1984)\cite{Hillas84}
identified possible sites of acceleration of UHECR based on
whether the putative source could contain the gyroradius of the
accelerated particles, and on the likely velocity of scattering
centres in these sites.  Following Hillas (1984)\cite{Hillas84} one
finds that possible sites included neutron stars
($10^7$--$10^{13}$G), gamma ray bursts and active galactic nuclei
($10^3$--$10^{4}$G), and lobes of giant radio galaxies and galaxy
clusters ($10^{-7}$--$10^{-5}$G).  This identification of
possible sources does not take account of energy losses
(synchrotron) and interactions (Bethe-Heitler and pion
photoproduction) which can cut off the spectrum, and so apply an
additional constraint which we discuss below (in his original
paper Hillas (1984)\cite{Hillas84} also used this additional constraint
to narrow the field of possible sources).

\begin{figure}[htb!]
\centerline{~~~~~~~~\epsfig{file=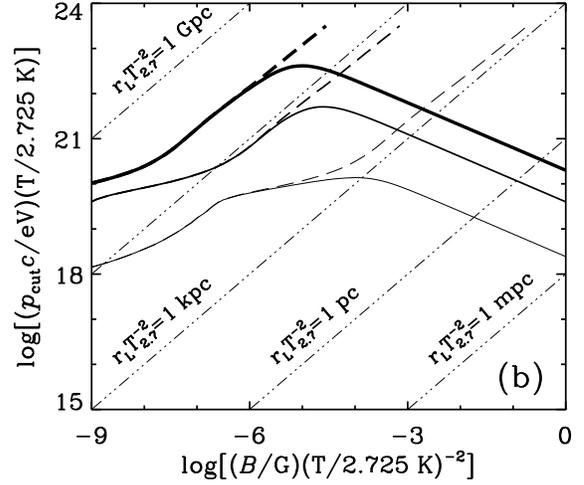,width=0.63\textwidth}}
\caption{Maximum energy as a function of magnetic field of
protons for maximum possible acceleration rate $\xi=1$ (upper
solid curve), $\xi=0.04$ (middle solid curve), $\xi=1.5 \times
10^{-4}$ (lower solid curve).  Dashed curves are limits from
Bethe-Heitler pair production and pion photoproduction only
(solid curves include synchrotron loss).  Dot-dot-dot-dash curves
are lines of constant Larmor radius as labelled. (From
Protheroe~2004\protect\cite{Protheroe2004}.)}
\label{photoprod_rate}
\end{figure}

For particle acceleration by electric fields induced by the
motion of magnetic fields $B$ (including those at astrophysical
shocks), the maximum rate of momentum gain by relativistic
particles of charge $Ze$ can be written (in SI units) $({dp /
dt})_{\rm acc} = \xi(p) Ze c B$ where $\xi(p) < 1$ is the
acceleration rate parameter and depends on the details of the
acceleration mechanism (see the review by Jones \& Ellison
\cite{JonesEllison91}1991, on the plasma physics of shock
acceleration, which also includes a brief historical review and
refers to early work).  To estimate cut-off momenta (or
energies), one needs the acceleration rate.  The following values
for the acceleration rate parameter have been suggested: maximum
possible acceleration rate $\xi(p_{\rm cut})$=1, plausible
acceleration at perpendicular shock with speed 0.1$c$,
$\xi(p_{\rm cut}) \approx 0.04$, and plausible acceleration at
parallel shock with speed 0.1$c$, $\xi(p_{\rm cut}) \approx
1.5\times$$10^{-4}$ \cite{Protheroe00}(Protheroe 2000).  Based on
the total momentum loss rate for Bethe-Heitler pair production
and pion photoproduction on the CMBR, synchrotron losses and
redshifting the proton cut-off momentum is plotted in
Fig.~\ref{photoprod_rate} as a function of magnetic field for
three adopted $\xi$-values (chain lines are for constant Larmor
radius as labelled).  This plot clearly shows that to accelerate
protons to $\sim 10^{20}$~eV large regions of relatively low
magnetic field $\sim 10^{-7}$--$10^{-3}$G are needed, apparently
ruling out high magnetic field regions for the origin of UHECR
(see also Medvedev~\cite{Medvedev03}2003).  One sees that, in
principle, protons can be accelerated up to $\sim 5 \times
10^{22}$~eV in Mpc scale region with $\sim 10^{-5}$G.

Returning to the Hillas plot (Fig.~\ref{proton_emax}),
constraints have been added corresponding to the three curves in
Fig.~\ref{photoprod_rate}, and the chain lines give constant
proton energy values as indicated.  Sources to the right of the
solid curves are excluded; a possible exception to this is in the
case of relativistically beamed sources (e.g.\ for AGN see
\cite{Protheroeetal03}Protheroe et al.\ 2003, and for GRB see
\cite{PelletierKersate00}Pelletier \& Kersate 2000) where
neutrons emitted along the direction of relativistic motion can
be Doppler boosted significantly in energy.  Another possible
exception is the case of so called ``one-shot'' mechanisms (e.g.\
\cite{Haswelletal92,Sorrell87}Haswell et al .\ 1992, Sorrell 1987)
where a particle is accelerated by an electric field along a
nearly straight path which is essentially parallel to the
magnetic field such that curvature and synchrotron losses are
negligible.  Suggested sites for this include polarization
electric fields arising in plasmoids injected into a neutron
star's magnetosphere (\cite{LitwinRosner01}Litwin \& Rosner 2001)
and magnetic re-connection in the magnetosphere of accretion
induced collapse pulsars (\cite{deGouvelaDalPinoLazarian01}de
Gouvela Dal Pino \& Lazarian 2001).  Another possibility is
plasma wakefield acceleration, i.e.\ acceleration by collective
plasma waves, possibly in the atmosphere of a GRB, or
``surf-riding'' in the approximately force-free fields of the
relativistic wind of a newly born magnetar (\cite{Arons03}Arons
2003).  In these cases it is unclear whether the requirements of
negligible radiation losses can be met.

Alternative scenarios for UHECR origin include emission and decay
of massive particles (``X-particles'') by topological defects
(TD) or decay of massive primordial particles. Because of the
resulting flat spectrum of particles (including neutrinos,
gamma-rays and protons) extending possibly up to GUT (grand unified
theory) scale energies, topological defect models have been
invoked to try to explain the UHECR.  Propagation of the spectra
of all particle species over cosmological distances is necessary
to predict the cosmic ray and gamma-ray spectra expected at
Earth.  In most cases this results in excessive gamma-ray fluxes at GeV
energies in addition to cosmic rays.  Massive relic particles on
the other hand, would cluster in galaxy halos, including that of
our Galaxy, and may give rise to anisotropic cosmic ray signals
at ultra high energies.  One possibility proposed for getting
around this is if an extragalactic source emits a very high
luminosity in UHE neutrinos, some of which interact with relic
neutrinos gravitationally bound to our galaxy producing
``Z-bursts'' which generate the events observed above the
expected GZK cut-off.  (see Protheroe \& Clay 2004
\cite{ProtheroeClay2004} for a recent review of UHECR.)

\subsection{Neutrino signatures}

One way of getting information about acceleration sources of
UHECR is through the spectral shape near acceleration cut-off.
One of the present authors (\cite{Protheroe2004}Protheroe 2004) has recently shown  that
in the case of protons the spectrum can actually be quite
sensitive to the astrophysical acceleration environment.  Despite
the fact that for extragalactic UHECR almost all spectral
information above the GZK cut-off is lost, significant
information is preserved in the spectrum of neutrinos produced as
a result of pion photoproduction interactions during propagation
(\cite{Protheroe2004}Protheroe 2004).  Furthermore, the spectrum of these GZK
neutrinos differs significantly from that in Z-burst and
topological defect (TD) scenarios, and of course the neutrinos
are not deflected by magnetic fields and so should point back to
where they were produced.  Hence, UHE neutrino astronomy will be
able provide much needed clues to the origin of the UHECR.

\begin{figure*}[htb!]
\centerline{\hspace*{12mm}\epsfig{file=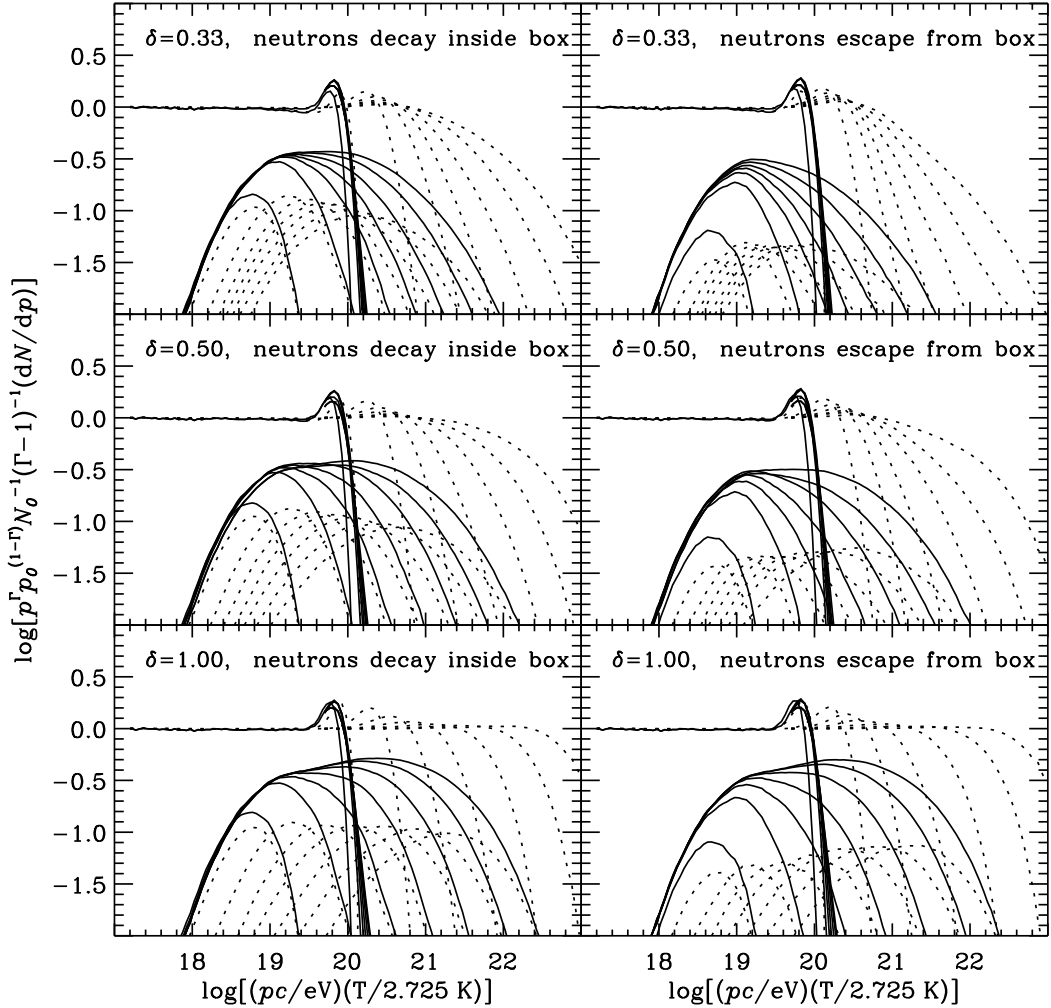,width=1.24\textwidth}}
\caption{Spectra of protons and neutrinos (all flavours) escaping
from the acceleration region (dotted curves) and after
propagation for time 100~Mpc/c (solid curves) for $p_{\rm cut}c= 10^{20}$
(leftmost curves), $10^{20.5}$, \dots , $10^{23}$~eV (rightmost
curves); $\Gamma=2$. (From Protheroe~2004\protect\cite{Protheroe2004}.)}
\label{evolv_photoprod_cutoff_mc}
\end{figure*}

The spectra of protons and neutrinos escaping from an
acceleration source and after propagation for 100~Mpc/c is shown
in Fig.~\ref{evolv_photoprod_cutoff_mc} for various possible
acceleration environments represented by the spectrum of magnetic
turbulence present (power-law dependence of acceleration rate),
the average magnetic field, its alignment, speed of scattering
centres (acceleration rate and maximum energy), and the size of
the acceleration region (decay or escape of photoproduced
neutrons).  As can be seen, while there is little difference in
the spectrum of UHECR after propagation over 100~Mpc, much
information is preserved in the spectrum of UHE neutrinos (``GZK
neutrinos'') produced during propagation as the UHECR flux is
eroded by the GZK-cutoff effect.  Of course, for very distant
sources UHECR would not be expected to be observed from
directions of sources.  Indeed, few if any may arrive at all
because of difficulty in reaching Earth through extragalactic
magnetic fields, whereas UHE neutrinos will arrive essentially
undeflected.  A very sensitive UHE neutrino telescope may
therefore observe neutrinos from extragalactic UHECR sources.
The diffuse GZK neutrino background can actually be quite large
if the UHECR sources evolve strongly with redshift (Engel et
al. 2001\cite{Engel01}).  Nevertheless, huge collecting areas
will be required for the detection of UHE neutrinos, and it is
here that the SKA through direct detection of neutrino-induced air
showers and, perhaps more importantly, through the detection of
Cherenkov radio transients from neutrino-induced showers in lunar
regolith may make a major contribution to understanding the
origin of the UHECR.

\section{Detection of high energy particles: historical background}
% my additions--PG, 04apr27-------------

There is clearly strong theoretical and phenomenological 
motivation to detect both the presumably hadronic
cosmic rays and associated neutrinos at EeV to ZeV energies.
The difficulty arises from the extremely low fluxes present--for
the highest energy cosmic rays at or above the $\sim 6 \times 10^{19}$~eV
GZK cutoff, one can expect of order a few per km$^2$ per century
at most. The associated EeV neutrino fluxes are, in the most
optimistic scenarios, perhaps 1-2 orders of magnitude larger than
this, but their detection efficiency is at most $\sim 1$\%
per cubic km of water-equivalent material, and thus the
neutrino rates are abysmally low in all existing and most 
planned detectors (though a recently approved NASA
long-duration balloon experiment,
ANITA~(\cite{ANITA}Barwick et al.\ 2003), may get an early, low-resolution 
view of these fluxes).

	\subsection{Giant air shower detectors.}  Since the early
1960's through the mid-1980's the highest energy cosmic ray
detectors were exclusively large ground arrays of scintillators
or Cherenkov counters making direct detection of secondary
particles, mainly electrons, gamma-rays, and muons within the
confines of the air shower itself as it impacts the ground (see
Nagano and Watson 2000\cite{NaganoWatson2000} for a review and
references to major air shower detectors).  At the highest
energies, air shower detectors gain much of their collecting
aperture by capturing the edges of showers whose cores fall
outside their fiducial array boundaries, sometimes by many
hundreds of meters.  Thus the shower energy must be estimated by
parametric models for the particle density at the shower
periphery -- a technique which has undergone much evolution
throughout the history of giant air shower detection, and still
retains much controversy in the details of its application even
today.

In the mid-1980's the first Nitrogen air fluorescence (N2fl) detector,
the Fly's Eye, came online. Since that time, both the Fly's Eye
and the follow-on High Resolution Fly's Eye (HiRes) have
become competitive with the air shower ground arrays in
their detection efficiency and aperture, and HiRes now has
the largest exposure and data sample of any detector to date.

The N2fl technique is very different than that of ground
array detection, since the detectors do not require direct
intersection with any portion of the air shower particles,
but rather detect the secondary incoherent radiation from
de-excitation of Nitrogen heated by the passage of the shower.
Such emission may be seen by optical telescopes of several
m$^2$ aperture out to tens of km distance from the shower
itself, and thus a small installation of modest, low-optical
quality (e.g., searchlight-style) reflectors can, by viewing
a good fraction of the surrounding sky, create an effective
air shower collecting aperture of several thousand
km$^2$~sr. The only drawbacks to this technique are its
sensitivity to atmospheric attenuation in the near ultraviolet
(where the nitrogen emission lies), and its requirement for
complete darkness and clear weather. These constraints lead
to a low net long-term duty cycle of less than 10\%, compared
to the $\sim 100$\% duty cycle of a ground array.

   \subsection{Cosmic ray air shower radio detection}
   
Interest in radio techniques for giant air shower detection stemmed
originally from the suggestion by Askaryan~(1962)\cite{Askaryan1962}
that any electromagnetic cascade in a dielectric material (gas, liquid
or solid) should rapidly develop net negative charge asymmetry due to
electron scattering processes and positron annihilation.  The net
electronic charge excess was estimated to be $\sim 20-30\%$, and
Askaryan proposed that Cherenkov radiation at wavelengths larger than
the longitudinal dimensions of the shower ($\sim 1$~m in air, and
$\sim 1$~cm in liquids or solids) would be emitted {\em coherently},
yielding a {\em quadratic} scaling of received power with the shower
energy.  This latter property immediately suggests that radio emission
might dominate the secondary radiation at the highest energies.  We
defer discussion of this so-called {\em Askaryan effect} in solids to
a later section; however, its application to air showers was
immediately noticed and pursued.

   \subsubsection{History}
Radio emission from cosmic ray air showers was discovered for the
first time by Jelley and co-workers in 1965 at a frequency of 44
MHz. They used an array of dipole antennas in coincidence with Geiger
counters. The results were soon verified and emission from 2 MHz up to
520 MHz was found in a flurry of activities in the late 1960's. These 
activities ceased almost completely in the subsequent years due to several
reasons: difficulty with radio interference, uncertainty about the
interpretation of experimental results, and the success of other
techniques for air shower measurements.

The radio properties of air showers are summarized in an excellent and
extensive review by Allan  (1971)\cite{Allan1971}. The main result of this review can
be summarized by an approximate formula relating the received voltage
of air showers to various parameters, where we also include the
presumed frequency scaling:

\begin{eqnarray}\label{crvoltage}
\epsilon_\nu &=& 20\, ~\mu {\rm V} {\rm ~m}^{-1} {\rm ~MHz}^{-1}
\left({E_{\rm p}\over 10^{17} ~{\rm eV}}\right)\sin\alpha ~~\times \nonumber \\
&~~& \,\cos\theta\,
\exp \left({-R\over R_0(\nu,\theta)}\right) \left({\nu\over55 ~{\rm
MHz}}\right)^{-1}.
\end{eqnarray}

Here $E_{\rm p}$ is the primary particle energy, $R$ is the offset
from the shower center and $R_0$ is around 110 m at 55 MHz, $\theta$
is the zenith angle, $\alpha$ is the angle of the shower axis with
respect to the geomagnetic field, and $\nu$ is the observing frequency
(see also Allan et al.~1970\cite{Allen_etal1970}; Hough \& Prescott
1970\cite{HoughPrescott1970}). The leading factor of 20 has been
disputed over the years since it was first published, and could be an
order of magnitude smaller.\footnote{More likely, the controversy over
this coefficient probably stems from the wide variation in measurement
conditions and the uncertainties in the flux calibration of the radio
antennas as well as in the energy calibration of the particles.}

The voltage of
the {\it unresolved} pulse in the coherent regime ($\nu \leq 100~$MHz)
can be converted into an equivalent flux density (a flux density for a 
steady continuum source required to produce the same energy over the
bandwidth limited time interval $\Delta t$) in commonly used
radio astronomical units

\begin{eqnarray}
S_{\nu} &=& \epsilon_\nu^2 \epsilon_0 c/\Delta t ,
\end{eqnarray}
\begin{eqnarray}S_{\nu} &=& 27\,{\rm MJy}\; \times \nonumber \\ &&
\left({\epsilon_\nu\over 10 ~\mu {\rm V}~ {\rm m}^{-1} {\rm~
MHz}^{-1}}\right)^2 \left({\Delta t\over\mu s}\right)^{-1}.\label{fluxconv}
\end{eqnarray}

The pulse duration is $\Delta t\sim1/\Delta \nu$ if the measurement is
bandwidth-limited. Note, that for larger bandwidths and hence higher
time resolution the energy of the pulse itself does not increase,
however, the equivalent flux density of a steady source needs to
increase, in order to produce an energy comparable to the pulse in the
shorter time interval. In the earlier measurements the pulses were
always unresolved when observing with $\Delta
\nu\simeq1$ MHz.

The formula was determined experimentally from data in the
energy regime $10^{16}\;{\rm eV}\,<E_{\rm p}<10^{18}$~eV. The flux
density around 100 MHz seems to depend on primary particle energy as
$S_\nu\propto E_{\rm p}^2$ (Hough \& Prescott 1970\cite{HoughPrescott1970}; Vernov et
al. 1968\cite{Vernov_etal1968}; Fig.~\ref{Edependence}) as expected for coherent emission
(see below). This dependency is, however, not yet undoubtedly
established, since a few earlier measurements apparently found
somewhat flatter power-laws (Barker et al. 1967\cite{Barker_etal1967} as quoted in Allan \cite{Allan1971}
1971).

\begin{figure}[htb!]
\centerline{\psfig{figure=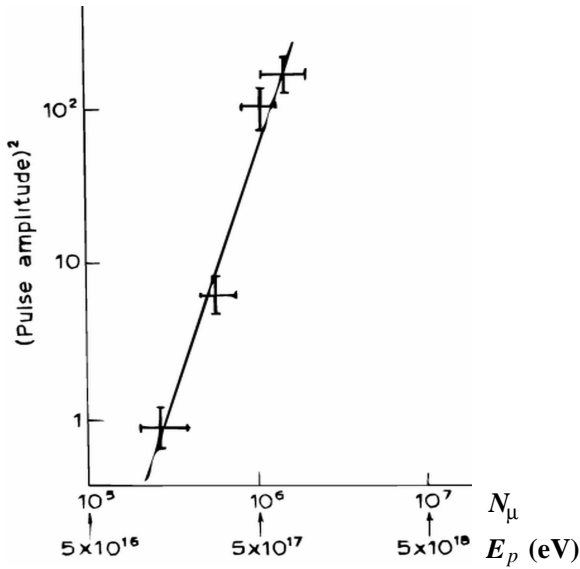,width=0.48\textwidth}}
\caption[]{\label{Edependence}The dependence of EAS radio flux on
the primary particle energy as measured by Vernov et al. (1968)\cite{Vernov_etal1968}
following roughly a $E_{\rm p}^2$ power-law. Some earlier papers found
somewhat flatter dependencies.}
\end{figure}

\begin{figure}[htb!]
\centerline{\psfig{figure=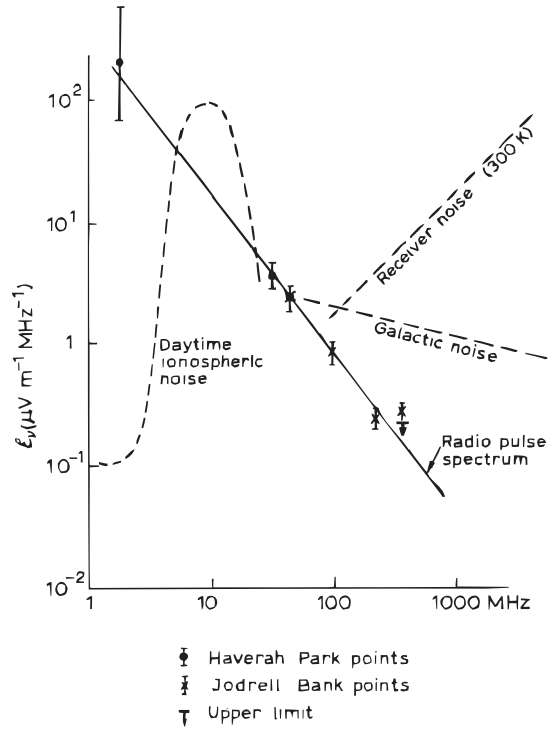,height=0.6\textwidth}}
\caption[]{\label{crspectrum}A tentative radio pulse spectrum for 2
MHz to 520 MHz. The data are not simultaneous. From Allan (1971)\cite{Allan1971} and
Spencer (1969)\cite{Spencer1969}.}
\end{figure}

Very little concrete data exist on the spectral dependence of EAS
(Extensive Air Shower) radio emission (e.g., Spencer
1969\cite{Spencer1969}). Figure \ref{crspectrum} shows a tentative EAS
radio spectrum with a $\nu^{-2}$ dependence for the flux density
($\nu^{-1}$ dependence for the voltage). The 2 MHz data point was made
with a different experiment and there is a possibility that the
spectrum is somewhat flatter between 10-100 MHz (see Datta et
al. 2000), but this is not verified. The polarisation of the emission
could be fairly high and is basically along the geomagnetic E-W
direction (Allan, Neat, \& Jones 1967\cite{AllanNeatJones1967}) which
strongly supports an emission mechanism related to the geomagnetic
field.

\begin{figure}[htb!]
\centerline{\psfig{figure=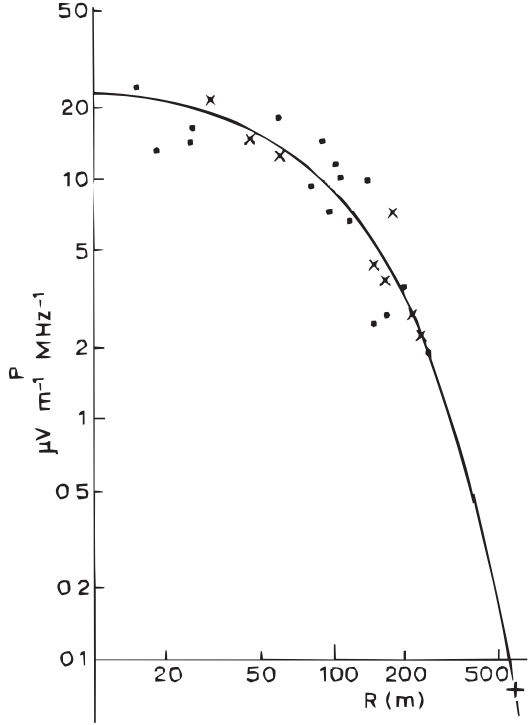,height=0.6\textwidth}}
\caption[]{\label{beamshape} Normalized radio pulse amplitudes in
$\mu$V m$^{-1}$ MHz$^{-1}$ at 55 MHz as a function of distance $R$ in
meters from the shower axis. Each data point corresponds to one
measured cosmic ray event. The amplitudes were normalized to a
reference energy of $E_{\rm p}=10^{17}$~eV assuming the above
mentioned linear dependence of voltage on primary particle energy. The
measurements were made for zenith angles $\theta<30^\circ$. Crosses
and dots represent different particle energy bins between $10^{17}$~eV
and $10^{18}$~eV. The plus sign at 500 meters marks a single $10^{19}$
eV event. From Allan (1971)\protect\cite{Allan1971}.}
\end{figure}

Finally, one needs to consider the spatial structure of the radio
pulse. The current data strongly supports the idea that the emission
is not isotropic but is highly beamed in the shower direction. Figure
\ref{beamshape} shows EAS radio pulse amplitude measurements as a
function of distance $R$ from the shower axis -- the flux density
drops quickly with offset from the center of the shower. The
characteristic radius of the beam is of order 100 meter for
a $10^{17}$~eV vertical shower, with the emission originating at
5-7 km distance above an observer at sea level. The implied 
angular diameter of the beam is thus $\Theta \simeq 0.2/6 = 1.9^{\circ}$.

   	\subsubsection{The synchrotron model and recent work.}
 
Experiments have clearly established that cosmic ray air showers
produce radio pulses. The original motivation was due to a suggestion
from Askaryan~(1962)\cite{Askaryan1962} who argued that annihilation of positrons would
lead to a negative charge excess in the shower, thus producing
Cherenkov radiation as it rushes through the atmosphere. At radio
frequencies the wavelength of the emission is larger than the size of
the emitting region and the emission should be coherent. The radio
flux would then grow quadratically with the number of particles rather than
linearly and thus would be greatly enhanced.  This effect is important in
dense media where it was already experimentally verified (Saltzberg et
al. 2001; see below) and is important for detecting radio emission
from neutrino showers in ice or on the moon.

However, the dependence of the emission on the geomagnetic field
detected in several later experiments indicates that another
process may be important.  The basic view in the late 60's was
that the continuously created electron-positron pairs were then
separated by the Lorentz force in the geomagnetic field which led
to a transverse current in the shower. If one considers a frame
moving along with the shower, one would observe electrons and
positrons drifting in opposite directions impelled by the
transverse electric field induced by the changing geomagnetic
flux swept out by the shower front.  (Only in the case of shower
velocity aligned with the magnetic field lines will this induced
electric field vanish).  This transverse current then produces
dipole (or Larmor) radiation in the frame of the shower. When
such radiation is Lorentz-transformed to the lab frame, the boost
then produces strongly forward-beamed radiation, compressed in
time into an electro-magnetic pulse (EMP).  This was calculated
by Kahn \& Lerche (1966)\cite{KahnLerche1966} and also Colgate
(1967)\cite{Colgate1967}.

Falcke \& Gorham (2003)\cite{FalckeGorham03} suggested it might be
better to think of the emission simply as being synchrotron-like in
the earth's magnetic field, or ``coherent geosynchrotron emission'',
as they called it. This process is probably equivalent to the previous
suggestions since it is derived from the basic formula for dipole
radiation and the Poynting vector but does not require a consideration
of charge separation: The different sign of the charges is canceled by
the opposite sign in the Lorentz force for electrons and pairs and
hence both contribute in exactly the same way to the total flux (radio
astronomers will surely remember that an electron/positron plasma
produces almost the same amount of synchrotron emission as a pure
electron plus proton plasma). 

The basic and intuitive derivation of this effect can be found in
Falcke \& Gorham (2003)\cite{FalckeGorham03} using standard
synchrotron radiation theory. One important effect which is explicitly
neglected by this simple treatment is the Fresnel zone problem --
vertical air showers at $10^{19}$~eV reach their particle maximum at
ground level, and the radio emission arrive nearly simultaneously to
the particle ``pancake,'' indicating that the far-field conditions,
where the radiation field has had time to become well-separated from
its source, are not satisfied. Any estimate of the details of the
received radio emission which is intended to help with detailed
detector design, such as what may be required to justify any impact on
SKA parameters or planning, must therefore treat the problem with much
greater fidelity.

Such high-fidelity simulations of geosynchrotron emission are now
beginning to appear in the literature, and as the interest in this
approach grows, along with the compelling nature of the ultra-high
energy cosmic ray problem, the simulations can be expected to improve
as well. In the following section, we describe recent results in this
direction.

\subsubsection{Air shower electrodynamics: detailed modeling and Monte Carlo simulation.}

The challenge of developing high-fidelity air shower radio simulations breaks into
three distinct problems:
\begin{enumerate}
\item The adaptation of existing air shower simulation codes to 
provide the particle identification and sampling needed for 
electrodynamics modeling;
\item The implementation of actual electrodynamics computation within
the modified air shower code, and the development of radiation propagation
model; and 
\item the modeling of the detector geometry and detection process.
\end{enumerate}

To date, no group has implemented all three aspects of this program, but we
describe here two efforts which have gone much further than others in
addressing the difficult problem of the electrodynamics and detection
modeling.

\paragraph{Simulations by the Chicago/Hawaii group.}
One result with a first-order electrodynamics Monte Carlo simulation has been
completed by Suprun~et al.~(2003)\cite{Suprun03} in a joint effort of the Univ. of Chicago group 
headed by Jon Rosner, along with one of the current authors (P. Gorham) of this chapter. This
study investigated a $10^{19}$~eV vertical air shower, including explicit geomagnetic
effects, with general interest in elucidating issues for detection by a possible
radio augmentation to the Auger Observatory for ultra-high energy cosmic rays.

The Suprun et al. simulation did not make any simplifying assumptions regarding
far-field conditions.
Instead, the electrodynamics simulation began with the general
formula for a radiating 
particle~(\cite{Jackson,ZasHalzenStanev92}Jackson 1999, Zas et al.\ 1992) in arbitrary motion: 
\begin{eqnarray}
{\bf E}({\bf x},t_a) &=&
\frac{e \mu}{4 \pi \epsilon_0 } \,\left[\frac{{\bf
n}-n\boldsymbol{\beta}} {\gamma^2 
|1-n\boldsymbol{\beta}\cdot{\bf n}|^3\,l^2}\right]_{\rm ret} +  \nonumber \\
&~&\frac{e \mu}{4 \pi \epsilon_0
c}\,\left[\frac{{\bf n}\times\left[({\bf
n}-n\boldsymbol{\beta})\times\dot{\boldsymbol{\beta}}\right]}
{|1-n\boldsymbol{\beta}\cdot{\bf n}|^3\,l}\right]_{\rm ret}
\label{elfield} 
\end{eqnarray}
which is correct regardless of the distance
to the antenna. In this formula $\boldsymbol{\beta}$ is the
velocity vector in units of $c$,
$\dot{\boldsymbol{\beta}}=d\boldsymbol{\beta}/dt$ is the
acceleration vector, divided by $c$, ${\bf n}$ is a unit vector
from the radiating particle to the antenna, and $l$ is the
distance to the particle. $\mu\approx1$ denotes the relative
magnetic permeability of air, $n$ the index of refraction. The
square brackets with subscript ``ret" indicate that the quantities in
the brackets are evaluated at the retarded time, 
not at the time $t_a$ when the signal arrives at the antenna.

The first term decreases with distance as $1/l^2$ and represents a
boosted Coulomb field. It does not produce any radiation. The
magnitudes of the two terms in Eq.~(\ref{elfield}) are related as
$1/(\gamma^2 l)$ and $|\dot{\boldsymbol{\beta}}|/c$.
The characteristic acceleration of a 30~MeV electron ($\gamma\approx60$) of
an air shower in the Earth's magnetic field
($B\approx0.5$~Gauss) is $|{\bf a}|=ecB/(\gamma m) \approx
4.4\cdot10^{13}$~m/s$^2$.
Even when an electron is as close to the antenna as 100~m,
the first term is two orders of magnitude
smaller than the second and can be neglected. The second term falls as $1/l$
and is associated with a radiation field. It describes the electric field of
a single radiating particle for most geometries relevant to extensive
air showers. It can be shown (\cite{FW}Wheeler \& Feynmann 1949) to be proportional to the apparent
angular acceleration of the charge up to some non-radiative terms that are
proportional to $1/l^2$. This relation is referred to in the literature as
``Feynman's formula.'' 

Suprun et al. did not, however, yet perform a full cascade
calculation, but rather used a parametrisation of the shower density to
generate a shower profile, then used Monte Carlo techniques to
sample the particle distribution obeying this parametrisation.
In one of the longest-standing empirical models for air shower
development, called the Nishima, Kamata, Greisen (NKG) model,
the lateral particle density $\rho_e$ is parametrized by the age parameter $s$ 
of the shower ($s=1$ for the shower maximum) and the Moli\`ere radius
$r_m$ (\cite{lateral,G,NK}Bourdeau et al.\ 1980, Greisen 1956, Kamata \& Nishimura 1958):
\beq
\rho_e = K_N\,\left(\frac{r}{s_m r_m}\right)^{s-2}\,
\left(1+\frac{r}{s_m r_m}\right)^{s-4.5},
\label{eqn:lateral}
\eeq
where
\beq
K_N=\frac{N}{2\pi s_m^2 r_m^2}\,
\frac{\Gamma(4.5-s)}{\Gamma(s)\Gamma(4.5-2s)}  ,
\eeq
$\Gamma$ is the gamma function, $r$ the distance from the shower axis, $N$
the total number of charged particles, and $s_m=0.78-0.21s$. 
The Moli\`ere
radius for air is approximately given by $r_m=74\,(\rho_0/\rho)$~m, with 
$\rho_0$ and $\rho$ being the air densities at sea level and the altitude 
under consideration, respectively.

As a shower travels toward the Earth and enters denser layers of the
atmosphere, the age parameter increases while the Moli\`ere radius drops. 
Both processes affect the spread of the
lateral distribution. The influence of the age parameter appears to be
more significant. As it grows, the average distance of the shower
particles from its axis increases. This effect overcomes the influence of a
smaller Moli\`ere radius which tends to make the lateral distribution more
concentrated toward the axis. For a fixed age parameter $s$, however,
the Moli\`ere radius is the only quantity
that determines the spread of the lateral distribution.
At shower maximum ($s=1$) the average distance from the axis can be
calculated to be $(2/3) s_m r_m = 0.38\, r_m$.

\begin{figure}[htb]
\centerline{\includegraphics[width=.48\textwidth]{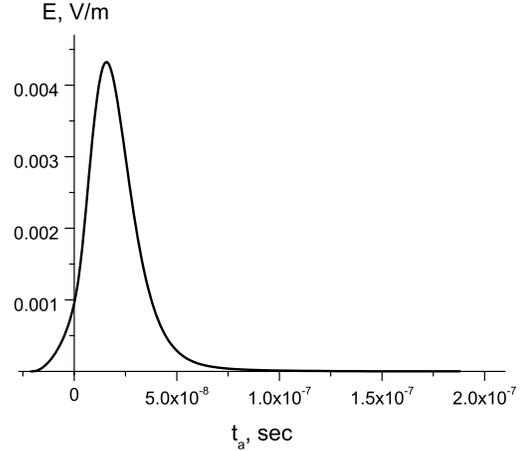}}
\caption{The EW component $E_{EW}$ of electromagnetic pulse of
$0.33\cdot10^{10}$ radiating
electron-positron pairs distributed over the thickness of the shower pancake
at 1800~m above sea level. The axis of the pancake is located 200~m South
of the antenna. The time axis was chosen in such a way that the pulse 
produced by a pair located in the axis at the bottom of the pancake starts 
at time 0.}
\label{fig:pancake}
\end{figure}

\begin{figure}[htb]
\centerline{\includegraphics[width=.48\textwidth]{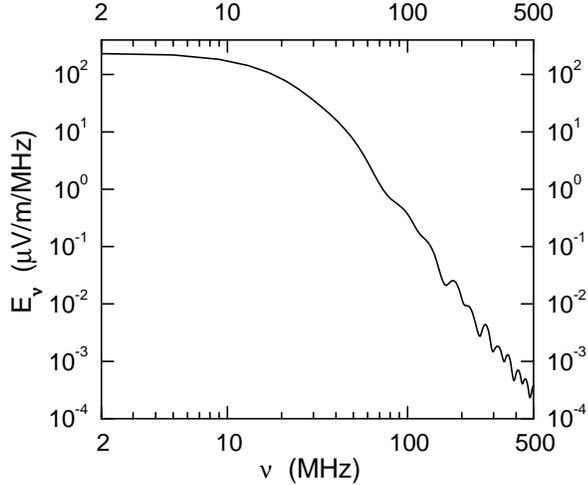}}
\caption{The Fourier transform of the electromagnetic pulse shown in
Fig.~\ref{fig:pancake}. The spectrum is very flat below 2~MHz.
The limited statistics of the model results in some jitters at 200$-$500~MHz.
The spectrum above 500~MHz is not shown because the statistics is 
not sufficient to make reliable predictions of the Fourier components at 
these high frequencies.}
\label{fig:FT}
\end{figure}

Fig.~\ref{fig:pancake} shows the results of the Suprun et al. simulation for
the shape of the intrinsic radio pulse, in terms of field strength vs.
time at the receiving antenna location, though without any of the filtering
effects of any antenna imposed on it yet. Fig.~\ref{fig:FT} gives the Fourier
transform $E_\nu$ of this pulse.  The nonzero thickness of the air-shower pancake
translates into a loss of coherence at frequencies corresponding to wavelengths
comparable to the shower thickness,
thereby limiting the main part of the radiation spectrum to the frequencies 
below 100~MHz. 

% This is Figure 5

These simulations, though using a greatly thinned set of input particles
($10^4$ compared to $10^{10}$ in actuality) do show characteristics similar
to what was observed historically. In addition, the simulations also begin
to reveal some of the geomagnetic complexity of the emission pattern,
suggesting reasons for some of the surprising variations observed in the
measurements by antenna arrays.

Consider the frame centered at the antenna, with axis $Ox$ going
to  the magnetic West, $Oy$ to the South and $Oz$ directly up. The
initial velocity of all charged particles is assumed to be
vertical: $\boldsymbol{\beta}=(0,0,-1)$, while the initial
acceleration $\dot{\boldsymbol{\beta}}$ is parallel to $Ox$, or,
in other words, to the $(1,0,0)$ vector.
%(see Fig.~\ref{fig:geometry})
 
Electrons bend toward the magnetic West
and positrons toward the East. The electric fields from both
particles of an electron-positron pair are coherent; the opposite
signs of their accelerations are canceled by the opposite signs
of the electric charges.

Let $\psi$ be the angle between
$Ox$ and the direction to the shower core, $R$ the distance to the
core, and $h$ the altitude of the radiating
particle above the antenna.
The denominator of the second term of Eq.~(\ref{elfield}) is independent
of $\psi$.
The numerator determines that, to leading
(second) order in $R/h$, the initial
electric field vector ${\bf E}$ received at the antenna lies in the
horizontal plane and is
parallel to $(\cos2\psi,\sin2\psi,0)$ (\cite{NIM}Green et al.\ 2003):
\beq
{\bf E} \ \| \ (\cos2\psi,\sin2\psi,0).
\label{eqn:Eparallel}
\eeq
The magnitude of the numerator is
independent of the angle $\psi$ up to terms of order $R^4/h^4$.
This result shows that although particles are accelerated by the
Earth's magnetic field in the EW direction regardless of angle $\psi$,
the radiation received at the antenna does not show preference for the EW
polarisation. Instead, it is directly related to the angle $\psi$.  As the
particle trajectory bends in the Earth's magnetic field and
the velocity deflects from the vertical direction, the
relation~(\ref{eqn:Eparallel}) between
the direction of the electric field vector and angle $\psi$ does not hold.
Nonetheless, it will be useful for understanding the angular dependence of
the electric field.

\begin{figure}[htb!]
\centerline{\includegraphics[width=0.48\textwidth]{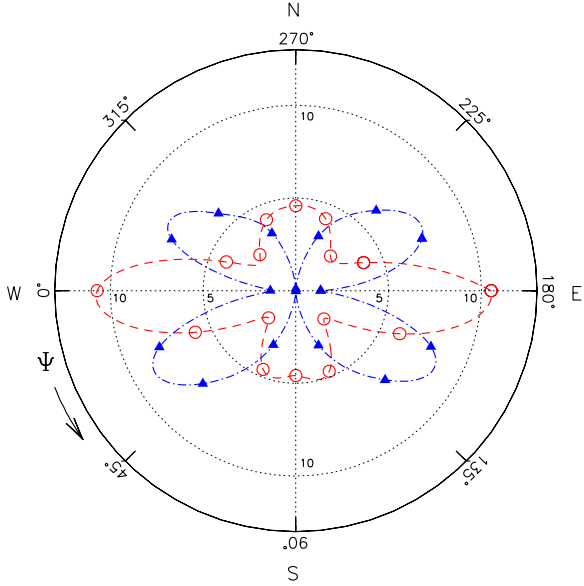}}
\caption{The East-West and North-South components of the field strength
$|{E}_{\nu EW}|$ and $|{E}_{\nu NS}|$ (circles and triangles,
respectively) at 55~MHz as functions of angle $\psi$ between the magnetic
West and direction to the shower core. The distance between the origin and a
circle or a triangle represents the field strength in the units of 
$\mu$V/m/MHz. The angular spacing between circles or triangles is 
$\pi/8$. At $\psi=\pm\pi/2$ $|{E}_{\nu NS}|$ do not 
exceed~0.1~$\mu$V/m/MHz and two triangles overlap.
All points were calculated for the vertical shower at a 200~m distance 
from the antenna.}
\label{fig:EWNS}
\end{figure}

Suprun et al.\ computed electromagnetic pulses for the pancakes with axes located at
the same distance $R=200$~m from the antenna but at various angles $\psi$
from the $Ox$ direction. Fig.~\ref{fig:EWNS} shows the radio
signal strengths that would be received by EW and NS-oriented antennas.
Note that Eq.~(\ref{eqn:Eparallel}) predicts that components of the radiation
coming from the start of the particle trajectory vanish at some angles 
$\psi$: $E_{EW}=0$ at
$\psi=\pm\pi/4,\,\pm3\pi/4$, while $E_{NS}=0$ at $\psi=0,\,\pm\pi/2,\,\pi$.
This fact explains why ${E}_{\nu EW}$ is relatively small at
$\psi=\pm\pi/4,\,\pm3\pi/4$ and ${E}_{\nu NS}$ is
small at $\psi=0,\,\pi$ (Fig.~\ref{fig:EWNS}).
Another mechanism is responsible for ${E}_{\nu NS}$ being virtually $0$ 
at $\psi=\pm\pi/2$. At these angles the trajectories of two charged particles
of an electron-positron pair are symmetric with respect to the $yOz$ plane. 
The NS component of radiation emitted by this pair
vanishes not only at the start but throughout its flight. 

% This is Figure 7

\paragraph{Modeling by the LOPES collaboration.}

Another simulation effort is under way in the Max-Planck-Institut für
Radioastronomie at Bonn, led by T. Huege and an author of this chapter
(H.\ Falcke). This group is part of a collaboration developing the
LOFAR Prototype Experimental Station (LOPES), an engineering model of
one station of the Low Frequency Array (LOFAR). LOPES is operating
jointly with the KASCADE Grande air shower array in Karlsruhe
(Schieler et al. 2003)\cite{Schieler2003}; LOFAR is a funded effort to
develop a very large area ground array for radio astronomy in the HF
to VHF regime, sharing many common interests with the SKA.  The LOPES
group has recently published a detailed analysis of the geosynchrotron
model for the case of a $10^{17}$~eV air shower
(\cite{HuegeFalcke02}\cite{HuegeFalcke03}Huege \& Falcke 2002, 2003)
in preparation for a major effort at an electro-dynamical air shower
Monte Carlo code (\cite{HuegeFalcke04}Huege \& Falcke 2004).

The LOPES group has taken special care of taking into account the
longitudinal development of the air shower by performing an
integration over the shower as a whole, and they have considered the
variation of the field strength as a function of radial distance from
the shower core as well. They use a shower parametrisation based on
the NKG model with a shower disk that flares out from the center, in a
manner similar to the Chicago/Hawaii study, and thus, apart from the
energy difference, the results do bear some comparison. The LOPES
study also did an integral over a power-law distribution of electron
energies, appropriate to an air shower. However, they did not do any
near-field corrections to their results, but this is not a major
drawback for a lower energy shower since these showers do reach their
maxima at altitudes of typically several km away from an observer on
the ground.

\begin{figure}[htb]
   \begin{center}
   \psfrag{nu0MHz}[c][B]{$\nu$~[MHz]}
   \psfrag{Eomega0muVpmpMHz}[c][t]{$\left|\vec{E}(\vec{R},\omega)\right|$~[$\mu$V~m$^{-1}$~MHz$^{-1}$]}
   \includegraphics[width=0.48\textwidth]{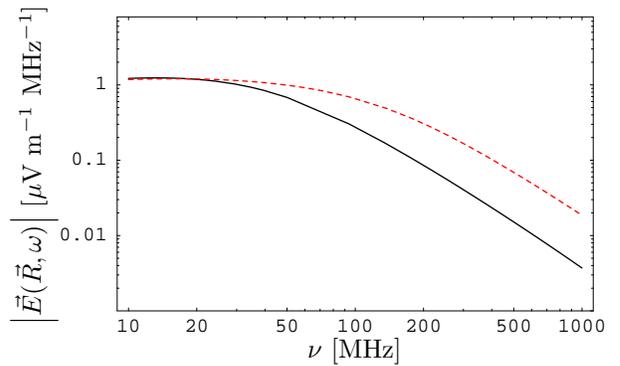}
   \caption{
   \label{fig:coherence_variants_flaring}
   $\left|\vec{E}(\vec{R},2\pi\nu)\right|$-spectrum at the centre 
   of the area illuminated by the maximum of a $10^{17}$~eV air 
   shower with flaring $\Gamma$-pdf, $R_{0}=4$~km and a broken 
   power-law energy distribution from $\gamma=5$--1000. Solid: 
   flaring \protect\cite{AgnettaAmbrosioAramo1997}(Agnetta et al.\ 1997) lateral 
   distribution, short-dashed: flaring \protect\cite{Linsley1986}(Linsley 1986) 
   lateral distribution
   }
   \end{center}
   \end{figure}

Fig.\ \ref{fig:coherence_variants_flaring} shows the spectrum
emitted by the air shower maximum for a shower disk profile with
realistic flaring according to the parametrisations of Agnetta et
al.\ (1997)\cite{AgnettaAmbrosioAramo1997} and Linsley
(1986)\cite{Linsley1986}.  As expected, the spectrum emitted by the
Linsley flaring disk extends to higher frequencies than the one
generated by the Agnetta flaring disk because of the lower
thickness in the shower centre where most of the particles
reside.
% (cf.\ Fig.\ \ref{fig:curvature_plot})

%______________________________________________________________
%______________________________________________________________

\begin{figure}[htb] \begin{center} \psfrag{R0m}[c][B]{distance
   from shower centre [m]}
   \psfrag{Eomega0muVpmpMHz}[c][t]{$\left|\vec{E}(\vec{R},\omega)\right|$~[$\mu$V~m$^{-1}$~MHz$^{-1}$]}
   \includegraphics[width=0.48\textwidth]{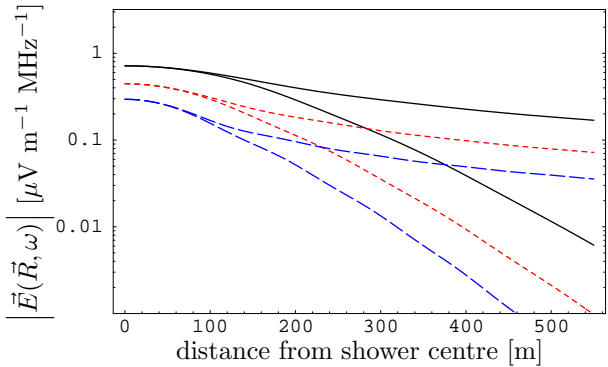} \caption{Radial
   dependence of $\left|\vec{E}(\vec{R},2\pi\nu)\right|$ for the
   maximum of a $10^{17}$~eV air shower with flaring (Agnetta et al.\ 1997)
   \protect\cite{AgnettaAmbrosioAramo1997} $\Gamma$-pdf,
   $R_{0}=4$~km and a broken power-law energy distribution from
   $\gamma=5$--1000.  Solid: $\nu=50$~Mhz, short-dashed:
   $\nu=75$~Mhz, long-dashed: $\nu=100$~Mhz, upper curves
   for distance from shower center to the east-west, lower curves for distance to north-south.
   \label{fig:radius_dependence_flaring}} \end{center}
   \end{figure}

The modeled radial dependence at different frequencies is shown in
Figure \ref{fig:radius_dependence_flaring}. Here the three families of
curves represent different frequencies, and the different slopes
between the two curves at a given frequency are for the cases of an
observer with a given distance from the shower center in the
directions perpendicular and parallel to the geomagnetic field. This
result thus indicates again the importance of the geomagnetic effects
in the azimuthal distribution of radiation for a given magnetic field
direction. Early results from the upcoming detailed Monte Carlo
simulations of the LOPES collaboration, however, show that asymmetries
in the emission pattern due to the geomagnetic field seem to be washed
out to a high degree once realistic distributions of particle track
lengths are taken into account.

%______________________________________________________________
%______________________________________________________________

%______________________________________________________________
   \begin{figure}[htb]
   \begin{center}
   \psfrag{time0ns}[c][B]{t~[ns]}
   \psfrag{Efield0muV0m}[c][t]{$\left|\vec{E}(t)\right|$~[$\mu$V m$^{-1}$]}
   \includegraphics[width=0.48\textwidth]{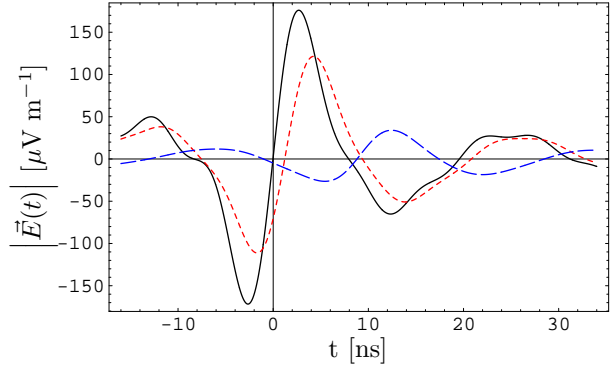}
   \caption{Reconstructed pulses emitted by the maximum of a $10^{17}$~eV 
   shower with flaring \protect\cite{AgnettaAmbrosioAramo1997} (Agnetta et al.\ 1997) $\Gamma$-pdf, 
   broken power-law energy distribution from $\gamma=5$--1000 and $R_{0}=4$~km, 
   using an idealized rectangle filter spanning 40--160~MHz. Solid: centre of illuminated area, 
   short-dashed: 100~m to north from centre, dash-dotted: 250~m to north from centre
   \label{fig:pulsesflaring}}
   \end{center}
   \end{figure}
%______________________________________________________________

Fig.\ \ref{fig:pulsesflaring} shows a reconstructed pulse generated 
by the flaring Agnetta disk as it would be measured by a receiver with 
a given bandwidth. The pulse amplitude 
drops noticeably when the observer moves from the centre of the 
illuminated area on the ground to a distance of 100~m, and is already 
quite diminished at a distance of 250~m. 

The LOPES study addresses the important problem of integrating over
the shower evolution as a whole in a simplified fashion by
approximating the shower evolution with a number of discrete
steps. The characteristic scale for these steps is given by the
``radiation length'' of the electromagnetic cascades in air,
$X_{0}=36.7$~g~cm$^{-2}$, corresponding to $\approx 450$~m at a height
of 4~km. One can therefore discretise the shower evolution into
``slices'' of thickness $X_{0}$, assuming these contain independent
generations of particles and therefore radiate independently.
Superposition of the individual slice emissions, correctly taking into
account the phases arising from arrival time differences, then leads
to the total emission of the shower.

For a vertical $10^{17}$~eV air shower at a height of $R_{0}=4$~km
they add the emission from eight slices above and eight slices below
the shower maximum to the emission from the maximum itself. The
closest slice then lies at $R_{0}=950$~m from the observer, a distance
they did not want to fall below because of approximations contained in
their calculations that are only valid in the far-field.

\begin{figure}[htb]
   \begin{center}
   \psfrag{nu0MHz}[c][B]{$\nu$~[MHz]}
   \psfrag{Eomega0muVpmpMHz}[c][t]{$\left|\vec{E}(\vec{R},\omega)\right|$~[$\mu$V~m$^{-1}$~MHz$^{-1}$]}
   \psfrag{Enu0muVpmpMHz}[c][b]{$\epsilon_{\nu}$~[$\mu$V~m$^{-1}$~MHz$^{-1}$]}
   \includegraphics[width=0.48\textwidth]{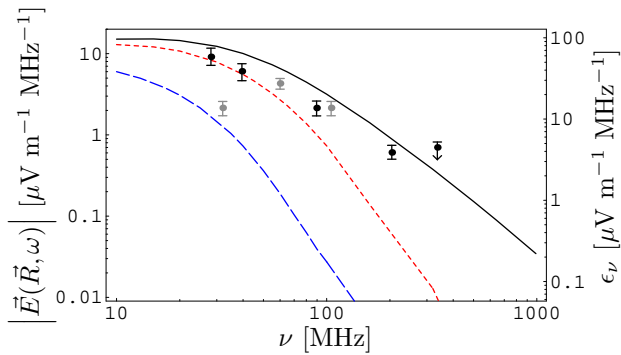}
   \caption{
   \label{fig:spectra_data}
   $\left|\vec{E}(\vec{R},2\pi\nu)\right|$-spectrum of a full (longitudinally integrated) $10^{17}$~eV 
   air shower with flaring \protect\cite{AgnettaAmbrosioAramo1997}  (Agnetta et al.\ 1997)
   $\Gamma$-pdf, $R_{0}=4$~km 
   and a broken power-law energy distribution from $\gamma=5$--1000. 
   Solid: centre of illuminated area, short-dashed: 100~m to north from centre, 
   long-dashed: 250~m to north from centre, black points: {\em{re-scaled}} 
   \protect\cite{Spencer1969}(Spencer 1969) data as presented by \protect\cite{Allan1971}(Allan 1971),
    grey points: {\em{re-scaled}} \protect\cite{Prah1971}(Prah 1971) data
   }
   \end{center}
   \end{figure}

Although this treatment is clearly oversimplified, the results
depicted in Fig.\ \ref{fig:spectra_data} indicate that the integration
over the shower as a whole significantly enhances the emission
strength and thus cannot be neglected. In particular, this implies
that the emission is actually not dominated by a narrow region around
the shower maximum, but that the entire shower evolution
contributes. A realistic treatment of the integration over the shower
as a whole is carried out as part of the upcoming Monte Carlo
simulations of the LOPES collaboration (\cite{HuegeFalcke04}Huege \&
Falcke 2004).

Data from the LOPES Experiment will become available soon, but first
results indeed confirm the association of the air shower with a sharp
radio pulse, having the expected properties (e.g., Horneffer et
al.~2004)\cite{Horneffer2004}. This puts the radio detection method on
rather firm ground.

%______________________________________________________________
%______________________________________________________________

It is interesting that in spite of the differences in the approach from the
LOPES studies and those of the Chicago/Hawaii group, the results for the
radio spectrum for a distance of 200/250 m from the shower core show a very similar
frequency dependence, with the field strength falling about a factor of 300 as one
goes from 10 to 100 MHz. The absolute value of the field strength is about
a factor of 30 or so different, which is inconsistent with a strict linear
scaling of field strength with energy as one might expect; however, agreement
to within a factor of 2-3 is actually quite good considering the fact that
these are completely independent efforts.

   \subsection{Askaryan effect and its confirmation}
As noted early in this discussion, the Askaryan effect was the original
motivation for much of the effort to measure radio emission from air
showers, but the coherent geo-synchrotron emission detailed above was found
to be the dominant contribution for air showers, and the coherent
Cherenkov emission from the charge excess, while not discounted, was largely
forgotten because of its small contribution. However, for showers in solid
materials such as ice or the lunar regolith which are relatively
radio-transparent, the shower lengths are short enough ($\sim 10$~m) that
the magnetic effects leading to synchrotron emission may be neglected, and
the coherent Cherenkov emission becomes the more important secondary radiation.
Here the quadratic rise of radio power with frequency leads to the conclusion
that, at energies above $10^{18}$~eV, the coherent Cherenkov emission will
dominate all secondary radiation, including optical emission, by a wide margin.
   
Although it is not presently possible to produce EeV cascades in terrestrial
accelerators, electromagnetic showers with composite total energies in this
range can be easily synthesized by super-posing gamma-rays of energies above
the pair-production threshold. If the gamma-ray bunch is small compared to
the wavelength of the radio emission (true for most pulsed linacs), the
resulting showers will differ from natural EeV showers only logarithmically,
due to the details of the initial interaction. However, since the bulk of the
radio emission arises from the region of maximum shower development, the
differences in radio Cherenkov emission are modest and easily quantified.

In mid-2000, Askaryan's hypothesis was in fact confirmed at the Stanford
Linear Accelerator Center (SLAC) in an experiment using a silica-sand
target and pulsed gamma-ray bunches with 
composite energies in the EeV range (Saltzberg et al.\ 2001)\cite{Saltzberg_01}.
In the 2002 follow-on experiment (Gorham et al.\ 2004)\cite{Gorham_04}, 
the sand was replaced by synthetic
rock salt, which has a higher dielectric constant and lower loss tangent
than silica sand, and further studies were made of the polarisation
behavior of the emission.

\begin{figure*}[htb!]
\centerline{\epsfig{file=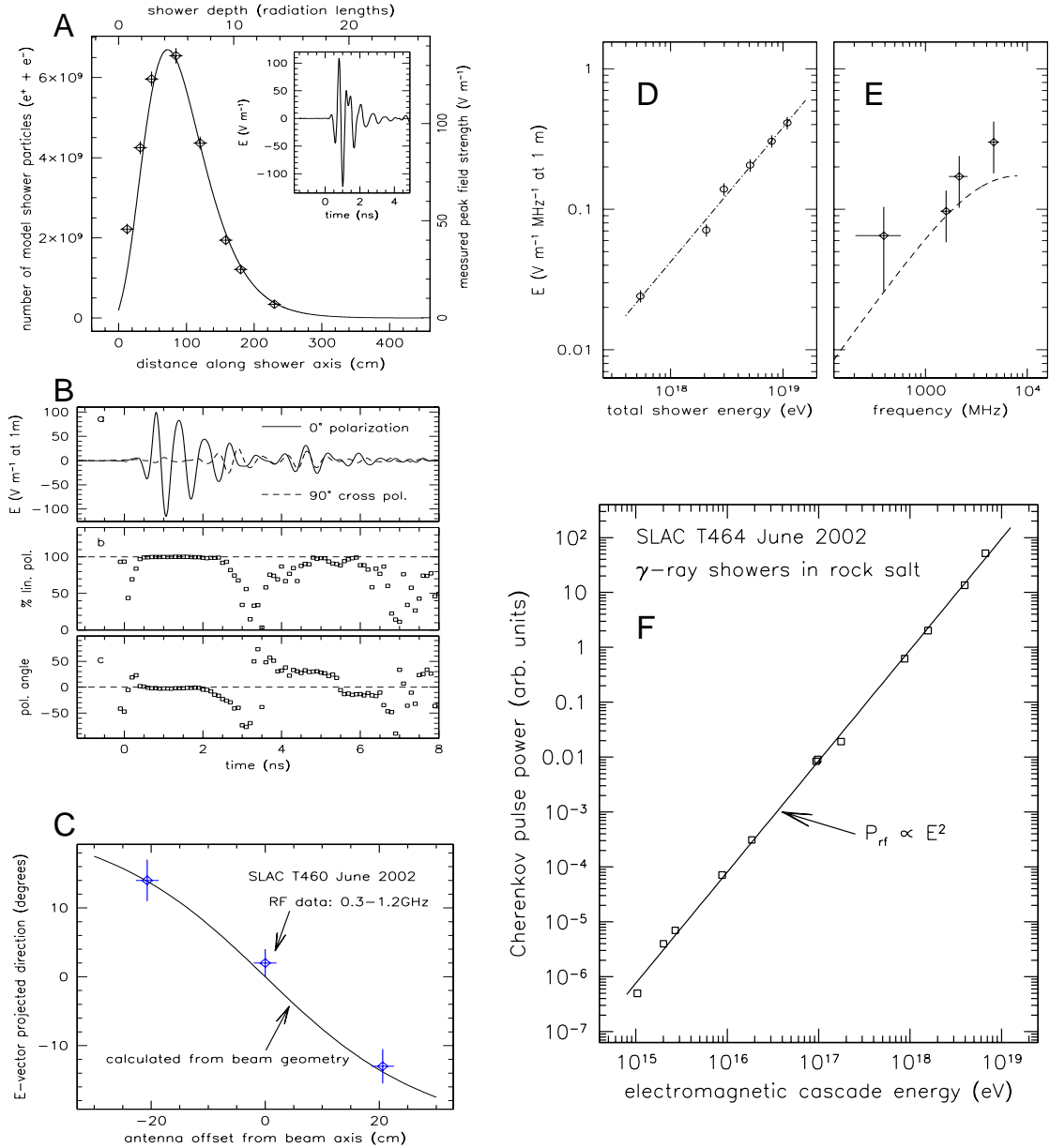,width=1.0\textwidth}}
\caption{ A: Shower RF field strength profile with typical pulse
(inset).  B: Polarisation measurements of a typical RF Cherenkov
pulse at 2 GHz.  C: Correlation of plane of polarisation with
antenna offset from shower axis.  D: Coherence of RF Cherenkov at
2 GHz, measured during 2000 SLAC experiment. E: absolute field
strength and prediction from Cherenkov.  F: Coherence of radiated
power over the 0.2-1.2 GHz band.
\label{Askaryan}}
\end{figure*}

Fig.~\ref{Askaryan}A shows a typical pulse profile (inset) and a set of measured
peak field strengths for pulses taken at different points along the shower
in the 2000 experiment.
The plotted curve shows the expected profile of the total number of particles
in the shower, based on the Kamata-Nishimura-Greisen\cite{Saltzberg_01}
approximation (Saltzberg 2001).
Here the field strengths have been scaled in the plot to
provide an approximate overlay to the relative shower profile. Clearly the
pulse strengths are highly correlated to the particle number profile. Since
the excess charge is also expected to closely follow the shower profile, this
result confirms Askaryan's hypothesis.

Pulse polarisation was measured with an S-band (2~GHz) horn directed at a
shower position 0.5 m past the shower maximum.  Fig.~\ref{Askaryan}B shows the
pulse profile for both the $0^{\circ}$ and $90^{\circ}$ (cross-polarised)
orientations of the horn. The lower two panes of this portion show the derived degree
of linear polarisation and the angle of the plane of polarisation,
respectively.  Because of the vector correlation of the pulse polarisation
with the shower velocity vector and the Poynting flux vector, it is
possible to use the angle of the polarisation to track the shower
axis. An example of this is shown in Fig.~\ref{Askaryan}C , where
the angle of the plane of polarisation is plotted at three locations with
respect to the shower axis, showing the high correlation with the
predicted angle.

%\begin{figure}[htb!]
%\epsfig{file=T464.eps,width=3.3in}
%\caption{\it Top left: Coherence of RF Cherenkov at 2 GHz, measured during 2000 SLAC
%experiment. Top right: absolute field strength and prediction from Cherenkov.
%Bottom: Coherence of radiated power over the 0.2-1.2 GHz band. \label{t464}}
%\end{figure}

Fig.~\ref{Askaryan}D shows a typical sequence of pulse field strengths
versus the total shower energy. The fitted linear
rise of field strength with beam current is consistent with complete
coherence of the radiation, implying the characteristic quadratic rise in the
corresponding pulse power with shower energy.
Fig.~\ref{Askaryan}F shows a similar result for the 2002 experiment, but now
covering a much wider range of energy, plotted as pulse power instead of
field strength. The Askaryan process is found to be quadratic over four
orders of magnitude in shower energy.

Fig.~\ref{Askaryan}E shows the spectral dependence of the radiation,
which is consistent with the linear rise with frequency that is also
characteristic of Cherenkov radiation. Also shown is a curve based on a
parametrisation of Monte Carlo results (Zas et al.\ 1992)\cite{ZasHalzenStanev92}. The
uncertainties are estimates of the combined systematic and statistical
uncertainties.  Note that the figure compares absolute field strength
measurements to the predictions and the agreement is very good.

In summary, there is clear experimental evidence that Askaryan's
hypothesis is confirmed and that the predicted emission from high energy
cascades is present in the expected amounts. This lends strong support to
experiments designed to exploit this effect for high
energy neutrino and cosmic ray detection.

\section{Prospects for the SKA}

The ultra-fast transient radio events described here will either be
signal, or at some level, background for the SKA. With recent pulsar
studies extending to broader and broader bandwidths, and faster and
faster pulse transients, understanding of these events may become important in
verifying the detection of pulsar transient events, certainly a mainstay of SKA
scientific interest. Whether or not the SKA can be a competitive instrument
for the detection of the various types of events described here will
depend strongly on the final choice of design. However, 
without careful choices made at this early stage of the effort, it
will be much more likely that the design is ``pessimized''
rather than optimized for their detection.

   \subsection{Cosmic ray air shower detection}
   
From the description above, it is evident that air shower radio emission
in the $\geq 10^{18}$~eV energy regime has three important 
characteristics which will impact the design of
any radio array with intent to detect them:
\begin{itemize}
\item The optimal frequency range is $\sim 20-200$~MHz;
\item The physical area over which one expects to detect the emission
is limited in diameter to a few tens of km, and often much less.
\item The time-scale for the air shower radio emission is an
impulse of order 20~ns or less in duration.
\end{itemize}
An SKA design extending down to 200~MHz is still adequate, though not
optimized, for air shower detection, but the sparsity of the array
and the bandwidth of the front-end receivers
will have significant impact on the triggering and reconstruction of the shower
energy and direction, and of course the ultimate sensitivity.
For the SKA to be competitive for giant air shower detection, 
careful consideration of all of these factors will make the
difference as to whether the SKA is irrelevant for this field, or
a dominant competitor.

Because of the wide variety of designs currently under consideration
for the SKA, and their rapid evolution, it is impractical to assess
each one for its capability in air shower radio detection. Instead,
we take the approach of estimating what would make the ideal
detector for air shower radio detection, and then we consider how
this compares to current plans.

As noted above, one would ideally like to work at frequencies that
extend below 100~MHz, but with as broad a bandwidth as possible, able
even to resolve the $\sim 10$~ns time-scale for air shower radio
emission.  This requires several hundred MHz of bandwidth, extending
perhaps down to 50~MHz or so. The immediate implication is that the
fundamental detector element must be a very broad-band antenna,
perhaps with of order 6:1 bandwidth or more. A high-gain antenna is
also problematic, since one cannot predict {\it a priori} the arrival
direction of an air shower radio pulse. For that reason a phased-array
concept with digital beam-forming is by far superior to other designs.
Dual polarisation is also clearly desirable, since the radiation
itself is highly polarised.

These considerations lead one to consider scale-invariant designs for
the primary antenna element such as dual-polarisation log-periodic
dipole antennas (LPDAs), but their typical beam-widths
($60-90^{\circ}$) would require a cluster of at least 6 antennas to
get even minimal coverage of the entire sky.  LPDAs have one other
characteristic which is undesirable for air shower detection: their
inherent pulse dispersion reduces sensitivity to impulsive events
unless a de-dispersing compensator either analog or digital) is
implemented. However, if this can be overcome, they are lightweight,
easy to construct and straightforward for impedance matching and
modeling. Variations on the LPDA design could also be scaled up as
stand-alone units to satisfy this need, and are more compact with
potentially better phase centers.  A non-dispersive alternative could
be an array of quad-ridged horns, which can routinely achieve the 6:1
bandwidth required, but they would be larger and heavier than
LPDAs. Obviously, even a simple active ``inverted-V'' antenna as used for
LOFAR and LOPES are also very useful if optimized for the right
frequencies.

The use of such a broad-band system of course raises the question of
how one can possibly deal with interference. This has been
successfully demonstrated with the LOPES experiment (see Horneffer et
al. 2003)\cite{Horneffer2004}. Another excellent example of a solution to this problem is
the FORTE satellite \cite{FORTE03}(Lehtinen et al.\ 2004), which
launched in 1997 with a 30-300~MHz nadir-pointing dual-polarisation
LPDA with a tunable 25~MHz receiving band. FORTE was optimized for
detection of electromagnetic impulsive events, and its mission was to
provide an unclassified test-bed for nuclear treaty verification
efforts while pursuing a science program of lightning and atmospherics
detection.

At an orbital altitude of 800~km,
FORTE was constantly exposed to a barrage of anthropogenic EM interference.
FORTE was able to retain triggering capability for impulses down to a level
within about $5\sigma$ of the ambient thermal noise level by 
sub-dividing their large band into a series of 1~MHz channels and
triggering when a majority of the bands exceeded threshold indicating
a broad-band pulse. The signal digitization was still done over the
entire bandwidth, preserving the broad-band coherence of the impulse.
But since the vast majority of anthropogenic interference
is inherently narrow-band, the multi-band trigger 
technique was very effective, when
combined with a so called {\it noise-riding} threshold which 
effectively maintained the trigger rate for each sub-band to a constant
level. This greatly reduced the ability of strong narrow-band carriers to
cause rapid re-triggering of one of the channels which might skew the
broad-band trigger rate. As a result, analysis of FORTE data has
recently even provided the first published limits on neutrino 
fluxes in energy regimes of $\sim 10^{22-24}$~eV, based on the lack
of observed radio impulses emanating 
from within the Greenland ice sheet \cite{FORTE03}(Lehtinen et al.\ 2004).

Applied to a potentially much broader-band system as proposed 
above for SKA air shower radio detection, the multi-band triggering
would in principle be applied to each cluster locally. If a trigger
occurred, it would cause a global broadcast out to stations 
within a several km radius of the triggered cluster,
interrogating these other stations to see if they also triggered.,
When enough stations trigger to justify it, a {\em global trigger} would
be initiated and all of the stations within the affected distance
(including appropriate margin to establish the boundaries of
the affected area) would save their buffered data.

The design implications for such a system clearly favor the phase
aperture array concept for the SKA, with a low frequency cutoff
at the lower end of the VHF band. In many ways the concept of a
Low Frequency Array (LOFAR) is perhaps best matched to air shower
radio detection, and insight can be gained toward adaptation of
the SKA possibilities by considering the adaptation required for
a LOFAR-type array.  Studies for such applications with LOFAR
have been recently published (Falcke \& Gorham 2003, Huege \&
Falcke 2002,
2003)\cite{FalckeGorham03,HuegeFalcke02,HuegeFalcke03}, and the
results are quite promising.

   \subsection{Neutrino detection}
In contrast to the problem of air shower radio detection with
the SKA, which is driven by the fact that there is only
one clear mechanism for detection, neutrino detection 
with the SKA may be pursued on several fronts. The scientific
motivations for both neutrino and air shower detection from
EeV to ZeV energies are closely related, and neutrino detection
at these energies will provide highly complementary information
to our current incomplete knowledge of the sources and
propagation of the highest energy cosmic rays.

To date, no cosmic high energy ($\geq 1$~GeV) neutrinos have
been detected from any source. The AMANDA detector at
Amundsen station, Antarctica, has detected cosmic-ray secondary
neutrinos up to $\sim 100$~TeV energies, but these arise from
interactions of garden-variety $\sim$~PeV cosmic rays in the
Earth's atmosphere.

For this reason, the discussion of neutrino detection must be
more broad in scope, since we do not yet know which detection channels 
might lead to methods with sufficient sensitivity to see fluxes
of neutrinos over the entire range of $10^{10-23}$~eV where they are
expected but so far unobserved. This section is therefore more
speculative with regard to possible techniques, but appropriate
to the high level of scientific interest in neutrino detection.

\subsubsection{Neutrino interactions in the Earth}

At energies of about 1 PeV, the earth becomes opaque to neutrinos at
the nadir. For higher energies, the angular region of opacity
grows from around the nadir till at EeV energies, neutrinos can
only arrive from within a few degrees below the horizon. The interaction
length at these energies is of order 1000 km in water, so such
neutrinos have a significant probability of interacting along a
$\sim 100$ km chord. If the interaction takes place within $\sim 10$~m
of the surface of rock or dry sand or soil, the resulting cascade will
produce coherent Cherenkov radiation up to microwave frequencies. Thus,
for example, since arrays are often sited with mountains or ridges near
the horizon, the entire near-surface volume of the mountain range
becomes a neutrino target, and events can originate anywhere along
its surface. The flux density
expected for such events (cf. Saltzberg et al. 2001) is
\begin{eqnarray}
S_{\nu} & \simeq &  12 {\rm~MJy}  
\left ( {R \over {\rm 1~km}} \right )^{-2} \times \nonumber \\ && \;\;\;\;
\left ( {E_{c} \over 10^{18}{~\rm eV}} \right )
\left ( {\nu \over 200 {~\rm MHz}} \right )^2
\end{eqnarray}
where $E_c$ is the cascade energy and $R$ the distance to
the cascade.\footnote{Note that in this case the neutrino energy is not
necessarily equal to the cascade energy $E_c$, because for
the typical deep-inelastic scattering interactions that occur
for EeV  neutrinos, only about 20\% of the energy is put into
the cascade, while the balance is carried off by a lepton.
For electron neutrinos, the electron will rapidly interact
and add its energy to the shower, but for muon or tau neutrinos,
this lepton will generally escape undetected (although the
tau lepton will itself decay within a few tens of km at 1 EeV).}
The Cherenkov process
weights these events strongly toward the higher frequencies,
though events that originate deeper in the ground will
have their spectrum flattened by the typical $\nu^{-1}$ behavior of
the loss tangent of the material.

A similar process leads to
coherent transition radiation (TR; cf. Takahashi et al. 1994\cite{Takahashi_etal1994}) 
from the charge
excess of the shower, if the cascade breaks through the
local surface. TR has spectral properties that make it more
favorable for an array at lower frequencies:
it produces equal power per unit bandwidth across the
coherence region.
The resulting flux density for a neutrino cascade breaking the
surface near the array array, observed at an angle of 
within $\sim 10^{\circ}$
from the cascade axis, is (cf. Gorham et al. 2000):
\begin{eqnarray}
S_{\nu,TR} (\theta \leq 10^{\circ}) &~\simeq~& 2~{\rm MJy} 
\left ( {R \over {\rm 1~km}} \right )^{-2} \times \nonumber \\ && \;\;\;\;
\left ( {E_c \over 10^{18} ~{\rm eV}} \right )^{2}~.
\end{eqnarray}
The implication here is that, if an array can retain some
response from the antennas to near-horizon fluxes, the
payoff may be a significant sensitivity to neutrino events
in an energy regime of great interest around 1 EeV,
or even significantly below this energy depending on the
method of triggering.

\subsubsection{Neutrino interactions in the atmosphere}

Neutrinos can themselves also produce air showers. The primary
difference between these and cosmic-ray-induced air showers
is that their origin, or first-interaction point, can be 
anywhere in the air column, with an equal probability of
interaction at any column depth. Neutrino air showers can
even be locally up-going at modest angles, subject to the earth-shadowing
effects mentioned above.

Detection of such events is identical to detection of cosmic-ray-induced
air showers, except for the fact that sensitivity to events
from near the horizon is desirable, since these will be most easily
distinguished from cosmic-ray-induced events. Beyond a zenith
angle of $\sim 70^{\circ}$ cosmic-ray radio events will be 
more rare, and those that are detected in radio will be distant. 
The column depth of
the atmosphere rises by a factor of 30 from zenith to horizon; thus
cosmic ray induced air showers have their maxima many kilometers away at
high zenith angles. Neutrino showers in contrast may appear close
by, even at large zenith angles.

Of particular interest is the possibility of observing ``double--bang''
(Learned \& Pakvasa 1995\cite{LearnedPakvasa1995}) 
tau neutrino events. In these events, a $\nu_\tau$ interacts first,
producing a near-horizontal air shower from a deep-inelastic 
hadronic scattering interaction. The tau lepton escapes with
of order 80\% of the neutrino energy, and then propagates an
average distance of $50 E_{\tau}/(10^{18}~{\rm eV})$~km 
before decaying and producing (in most cases) another shower
of comparable energy to the first. Detection of both cascades
within the boundaries of a surface radio array would provide a unique
signature of such events. And in light of the recent neutrino
results indicating $\nu_\mu \rightarrow \nu_{\tau}$ oscillations,
it is likely that neutrinos from astrophysically distant sources
would be maximally mixed, leading to a significant rate of
$\nu_{\tau}$ events.

   \subsubsection{Neutrino interactions in the lunar regolith}
   
There is an analogous process to the earth-surface layer cascades
mentioned above which can take place in the lunar
surface material (the regolith). In this case the cascade takes
place as the neutrino nears its exit point on the moon after having
traversed a chord through the lunar limb. This process, first
suggested by Dagkesamansky \& Zheleznykh (1989) \cite{Dagkesamansky}
is the basis of several searches
for diffuse neutrino fluxes at energies of $\sim 10^{20}$~eV
(Hankins et al. 1996\cite{Hankins}; Gorham et al. 1999\cite{Gorhametal1999}, 
2001\cite{GorhamRADHEP01}, Gorham et al.\ 2003\cite{GLUE}) using large
radio telescopes at microwave frequencies. Based on the 
simulations for these experiments (Alvarez M\~uniz \& Zas 1997a,
1997b\cite{AlvarezMunizZas1997,AlvarezMunizZas1998}; 
Zas, Halzen \& Stanev 1992\cite{ZasHalzenStanev92})
and confirmation through
several accelerator measurements (Gorham et al. 2000 \cite{Gorham_etal2000}; 
Saltzberg et al. 2001\cite{Saltzberg_01}), 
the expected flux density from
such an event at about 1 attenuation-length depth in the
regolith can be roughly estimated as
\begin{equation}
S_\nu~=~ 50\,{\rm Jy}\; \left ( {E_{c} \over 10^{20}{~\rm eV}} \right )
\left ( {\nu \over 200 {~\rm MHz}} \right )^2~.
\end{equation}
Note here that the flux density is far lower than for air shower events,
but the two should not be compared, since the lunar regolith events are
coherent over $\sim$~degree angular scales, corresponding to several
thousand km at the Earth's surface. They also originate from a small, known
angular region of the sky (the surface of the moon). Thus their detectability
depends on the sensitivity of the synthesized beam, and on the
ability of the system to trigger on band-limited pulses.

Transition radiation events may also be detectable in
a similar manner, as noted above. For TR from events that break the
lunar surface, the resulting pulse differs from a Cherenkov
pulse because it is flat-spectrum. Because TR is strongly
forward beamed compared to the Cherenkov radiation from the moon,
we estimate that the maximum
flux density for this case, at an angle of $\sim 1.5^{\circ}$
from the cascade axis, 
is about a factor of 20 higher than at $\sim 10^{\circ}$. At earth
the implied flux density for LOFAR is:
\begin{equation}
S_{max,TR} (\theta \simeq 1.5^{\circ}) ~\simeq~ 40\,{\rm Jy}\;
\left ( {E_c \over 10^{20} ~{\rm eV}} \right )^{2}~.
\end{equation}
Although this channel does not provide a higher flux density than
the Cherenkov process, it is a flat spectrum process that may 
in some cases provide
more integrated flux across a given band.

These pulses are essentially
completely band-limited prior to their entry into the ionosphere,
with intrinsic width of order 0.2 ns.
Dispersion delay in the ionosphere will of course significantly
impact the shape of any pulse of lunar origin. This will
limit the coherence bandwidth for a VHF system. The 
dominant quadratic part of the dispersion gives an overall
delay 
\begin{equation}
\tau_{ion} ~=~ 1.34 \times 10^{-7} {N_e \over \nu^2}
\end{equation}
where $\tau_{ion}$ is the delay in seconds at frequency $\nu$ (in Hz)
for ionospheric column density $N_e$ in electrons per m$^{2}$. For typical 
nighttime values of $N_e \sim 10^{17}$ m$^{-2}$
the zenith delay at 200 MHz is 330 ns, and the differential 
dispersion is of order 3 ns per MHz, increasing at lower frequencies as
$\nu^{-3}$. For bandwidths up to even several
tens of MHz for zenith observations, and perhaps a few MHz
at low elevations, the pulses should remain band-limited. However,
coherent de-dispersion will be necessary to accurately
reconstruct the broad-band pulse structure.

Although the problem of coherent de-dispersion is a difficult one, a
system operating in the 0.2-1~GHz range may have an edge 
in sensitivity over systems operating at
higher frequencies, under conditions where the intrinsic neutrino
spectra are very hard. This is due to the fact that the loss tangent
of the lunar surface material is relatively constant with frequency
(Olhoeft and Strangway 1976\cite{OlhoeftStrangway1976}), and thus the attenuation length
increases inversely with frequency.  This means that a lower frequency
array may probe a much larger effective volume of mass than the higher
frequencies can. At 200 MHz, the RF attenuation length should be of
order 50 m or more, compared to 5-7~m at 2 GHz.  When this larger
effective volume is coupled with the larger acceptance solid angle
afforded by the broader RF beam of the low-frequency Cherenkov
emission, the net improvement in neutrino aperture could well
compensate the loss of sensitivity at lower energies by a large
margin.

It is also worth noting here that these lunar regolith
observations are distinct from other methods in high energy
particle detection, in that they do require the array to track an
astronomical target, and can and will make use of the synthetic
beam of the entire array.  This is because, although the sub-array
elements should be used for the detection since they will have a
beam that covers the entire moon, the Cherenkov beam pattern from
an event of lunar origin covers an area of several thousand km
wide at earth, and is thus broad enough to trigger the entire
array. Post-analysis of such events can then localize them to a
few km at 100 MHz and about 200~m at 1.4 GHz on the surface of the
moon, providing opportunities for more detailed reconstruction of
the event geometry.  If, as expected, a high resolution (5~m) 3D
cartographic map of the moon will be produced by the Terrain
Mapping Camera of India's Chandrayaan-1 lunar mission due for
launch in 2008 [www.isro.org/chandrayaan-1/announcement.htm] becomes available,
this may be used to determine the local gradient and roughness of
the surface near the position of cascade exit.  We anticipate
that this information, together with the polarisation angle and
its frequency dependence, both of which can be measured by the
SKA, may enable neutrino direction reconstruction to an ``event
arc'' on the sky of thickness a few degrees to be routinely
achieved on an event-by-event basis.

\begin{figure}[htb!]
\centerline{~~~~~~~~~\epsfig{file=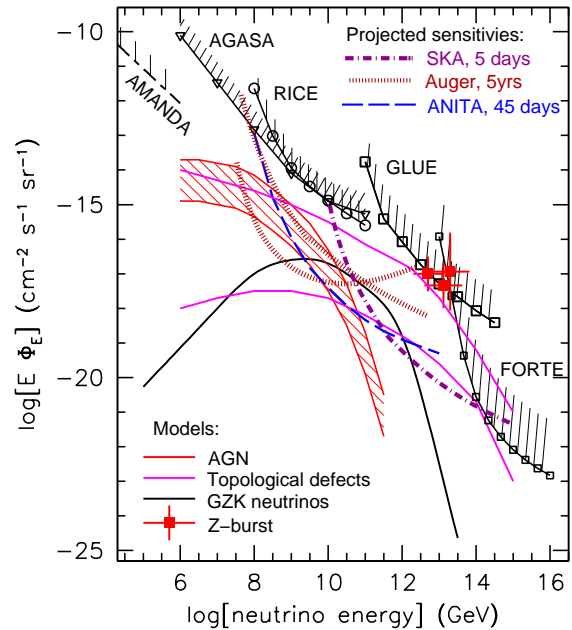,width=0.56\textwidth}}
\caption{Black solid curves with points show existing limits on diffuse
neutrino intensity (for references see \protect \cite{GLUE}Gorham et al.\ 2003,
from which this plot is adapted).  Thick, dark red dotted curves show the
expected sensitivity of Auger Observatory to $\nu_\mu$ and
$\nu_e$ (top), and $\nu_\tau$ (with no
deep-inelastic-scattering losses, bottom) in 5 years of observations.  
The magenta dash-dot curve labelled SKA gives an estimates
of the expected sensitivity of the SKA as a neutrino
observatories for 120~h of lunar observations. ANITA is
an Antarctic Long duration balloon mission due to fly in
2006; this estimate is for the full planned 3 flights by 2009.
Other curves
bracket AGN predictions, two TD models, and an estimate of the
maximal GZK neutrino flux.}
\label{nu_obslim}
\end{figure}

%\subsection{Summary \& outlook}

It is in the neutrino energy range most sensitive to UHECR origin at
$10^9$--$10^{13}$~GeV that the SKA may have the greatest impact by
using the lunar Cherenkov technique.  So far no UHE neutrinos have
been detected and the current observational limits are shown in
Fig.~\ref{nu_obslim}.

In the $10^{11}$--$10^{13}$~GeV range the Goldstone
Lunar Ultra-high Energy (GLUE) neutrino experiment \cite{GLUE}(Gorham et al.\ 2003)
has the best limit.  Other planned experiments in this energy
range such as SALSA \cite{SALSA}(Gorham et al.\ 2002) and 
ANITA \cite{ANITA}(Barwick et al.\ 2003) will
lower these limits and hopefully detect neutrinos.  The lunar
Cherenkov technique used in the GLUE experiment was pioneered by
Hankins, Ekers \& O'Sullivan (1996)\cite{Hankins} using the Parkes 64m
radio telescope.  The GLUE experiment used two dishes of the
Goldstone Deep Space Tracking Network for 120 hours to look for
Cherenkov radio emission from neutrino-induced cascades in lunar
regolith.  

By scaling relationships given by 
Gorham et al.\ (2000) and  Alvarez-Muniz \& Zas (2001) 
\cite{GorhamRADHEP01, MunizZas01, Beresnyak03} describing
the electric field strength at the radio telescopes expected for a
given cascade energy deposited in the regolith (see also Beresnyak
2003), and comparing the proposed technical specifications of the SKA
(assuming 1~GHz frequency will be used, its higher bandwidth and
larger telescope field of view and collecting area) with those of the
telescopes used in the GLUE experiment, one expects that for
comparable lunar observing time with the SKA the threshold will be
reduced to 2$\times$10$^{10}$~GeV and the sensitivity will be improved
by a factor of about 2000 (dot-dash curve ``SKA'' in
Fig.~\ref{nu_obslim}), making it potentially the most sensitive UHE
neutrino observatory in the future for covering a large part of the
important energy range $10^{9}$--$10^{14}$~GeV.

\section{Summary and outlook}

In the next several years giant air shower detectors will
investigate the spectrum, composition and anisotropy of the
UHECR, i.e.\ those with energies above $10^{10}$~GeV, in an
attempt to determine their origin.  Because of its large
collecting area, the use of the SKA to directly observe coherent
geosynchrotron radio emission from cosmic ray air showers has the
potential to make a significant impact in this field.  However,
cosmic rays are deflected by magnetic fields and do not point
back directly to their sources, and above $\sim
10^{11}$~GeV UHECR suffer severe energy losses on interacting with
CMBR photons, limiting their range to tens of Mpc from their
sources.  Hence studies of UHECR alone will probably be insufficient to
tie down their sources and whether they are accelerated or result
from the decay of massive relic particles or emission by topological defects.  

UHE neutrinos are the key to determining the origin of these UHECR.
This subject is of great importance to our understanding of the
Universe as it impacts on our knowledge of dark matter, gravity, and
high energy particle interactions.  Direct radio observation by the
SKA of air showers due to high energy neutrinos may contribute
significantly to high energy neutrino astrophysics, particularly below
$10^{10}$~GeV.  However, the enormous neutrino collecting area of the
Moon, together with the large aperture and excellent angular
resolution of the SKA make UHE neutrino astrophysics using the lunar
Cherenkov technique potentially the best approach for tying down the
origin of the very highest energy particles in nature.

The signal coincidence requirement between antennas and the
nanosecond duration signal experimental procedures are
significantly different from those in normal radio astronomy, and
must be taken into account together with the most appropriate
signal processing technique for multiple antennas in the design
of the SKA if it is to be used for lunar UHE neutrino
observations and take a leading role in neutrino astronomy at the
highest energies.

{\em Acknowledgement:} We thank T. Huege for useful comments on the document.

\end{document}